\documentclass[12pt]{article}
\usepackage{amsmath}
\usepackage{amssymb}
\usepackage{cancel}
\usepackage{graphicx}
\usepackage{wrapfig}

\renewcommand{\thefootnote}{\arabic{footnote}}

\begin{document}
\baselineskip=19.5pt

\begin{titlepage}

\begin{center}
\vspace*{17mm}

{\large\bf%
Heavy neutrino search in accelerator-based experiments
}

\vspace*{10mm}
Takehiko~Asaka,$^{*,\dag}$
%\footnote{~asaka@muse.sc.niigata-u.ac.jp}
~~Shintaro~Eijima,$^{*,\dag}$
%\footnote{~eijima@muse.sc.niigata-u.ac.jp}
~~Atsushi~Watanabe$^{\dag}$ 
%\footnote{~atsushi.watanabe@mpi-hd.mpg.de}
\vspace*{10mm}

$^*${\small {\it Department of Physics, 
Niigata University, Niigata 950-2181, Japan}}\\

$^\dag${\small {\it 
Max-Planck-Institut f\"ur Kernphysik, Saupfercheckweg 1, 69117 Heidelberg, Germany
}}\\

\vspace*{3mm}

{\small (December, 2012)}
\end{center}

\vspace*{7mm}

\begin{abstract}\noindent%
We explore the feasibility of detecting heavy neutrinos by the existing facilities
of neutrino experiments.
A heavy neutrino in the mass range $1\,{\rm MeV} \lesssim M_N \lesssim 500\,{\rm MeV}$
is produced by pion or kaon decay, and decays to charged particles which 
leave signals in neutrino detectors.
Taking the T2K experiment as a typical example, 
we estimate the heavy neutrino flux produced in the neutrino beam line.
Due to massive nature of the heavy neutrino, the spectrum of the heavy neutrino is  
significantly different from that of the ordinary neutrinos.
While the ordinary neutrinos are emitted to various directions in the laboratory
frame due to their tiny masses, the heavy neutrinos tend to be emitted to 
the forward directions and frequently hit the detector.
The sensitivity for the mixing parameters is studied by evaluating the number of 
signal events in the near detector ND280.
For the electron-type mixing, the sensitivity of T2K at $10^{21}$ POT is found to be 
better than that of the previous experiment PS191, which has placed the most stringent 
bounds on the mixing parameters of the heavy neutrinos for 
$140\,{\rm MeV} \lesssim M_N \lesssim 500\,{\rm MeV}$.
\end{abstract} 

\end{titlepage}

\newpage
\renewcommand{\thefootnote}{\fnsymbol{footnote}}
%%%%%%%%%%%%%%%%%%%%%%%%%%%%%%%%%%%%%%%%%%%%%%%%%%%%%%%%%%%%%%%%%%%%%%
\section{Introduction}
\label{intro}
%%%%%%%%%%%%%%%%%%%%%%%%%%%%%%%%%%%%%%%%%%%%%%%%%%%%%%%%%%%%%%%%%%%%%%
In the last few decades, neutrino oscillation experiments have 
conclusively shown that neutrinos are massive~\cite{nu}. 
The minimal version of the Standard Model is thus to be extended to
accommodate the neutrino masses.
In many possible extension of the Standard Model, neutral heavy leptons are
often predicted. 
In the seesaw mechanism~\cite{Seesaw} for example, the right-handed neutrinos are 
introduced and they weakly mix with the ordinary neutrinos after the electroweak 
symmetry breaking.

For the masses of the heavy neutrinos, a wide range of possibilities has been
discussed in literature.
In the canonical picture of the seesaw mechanism, heavy neutrino masses 
are supposed to be around the grand unification scale.
These super-heavy neutrinos can account for the baryon asymmetry of the universe
by the leptogenesis~\cite{leptogenesis}.
Another possibility to account for the baryon asymmetry has been suggested 
in~\cite{BAU,Asaka:2005pn}
and further studied in~\cite{BAU2}. In this scenario, two quasi-degenerate 
heavy neutrinos of $\mathcal{O}(100)\,{\rm MeV} - \mathcal{O}(10)\,{\rm GeV}$
play a crucial role in the early universe. 
Heavy neutrinos in the mass range $\sim 0.2\,{\rm GeV}$ could 
enhance the energy transport from the core to the stalled shock and favor the supernova 
explosion~\cite{Fuller:2009zz}.
Heavy neutrinos with a few keV mass have also attracted much interests as a 
viable dark matter candidate~\cite{DM} and an agent of the pulsar 
velocities~\cite{palsar}. 
Remarkably, the dark matter and the baryon asymmetry due to keV and GeV heavy neutrinos
can originate in a simple framework so called $\nu$MSM~\cite{Asaka:2005an,Asaka:2005pn}, 
which is an extension of the Standard Model with just three generations of the 
right-handed neutrinos.
Besides the super-heavy range much larger than TeV, such heavy neutrinos 
can be tested in existing and forthcoming experiments due to lower threshold energies 
of production (for example, see Ref.~\cite{Gorbunov:2007ak,Atre:2009rg} and 
references therein).

In this paper, we focus on the heavy neutrinos 
produced by kaon decays.
The previous neutrino experiments, including peak searches in the meson decays~\cite{pi,
pi2,pi3,K,K2}
and the decay searches with accelerators~\cite{acc,PS191}, have placed stringent bounds 
on the mixing parameters in this mass range. 
In particular, PS191~\cite{PS191} has placed the strongest bounds for the mixing
parameters in the mass range of $140\,{\rm MeV} \lesssim M_N \lesssim 500\,{\rm MeV}$.
Since the PS191 experiment in 1984, however, no further experiments
of this type of decay search have performed and the bounds have not been updated for
about 30 years.
On the other hand, great progress has been made in neutrino oscillation experiments 
over the same period. 
It is interesting to note that typical long or short baseline experiments are
equipped with the (near) detectors placed at $\mathcal{O}(100)$ meters away from
the beam targets, which detectors are capable of measuring the charged-particle tracks
produced by the heavy neutrino decays.
A natural question is then whether the existing accelerator-based neutrino
experiments are capable of discovering the heavy neutrinos and how sensitive
such experiments are\footnote{Far detectors with large volume are also useful to 
detect the heavy neutrinos produced in the
atmosphere~\cite{Kusenko:2004qc, ATM}.}.
We believe that this is a timely question to ask. 
In fact, the exposure of PS191 is about $200\,{\rm m^3} \times 
10^{20}\,{\rm POT}$ while the recent accelerator-based neutrino experiments
are expected to achieve $10^{21}\,{\rm POT}$ with the near detectors which are
typically no smaller than $200\,{\rm m^3}$ by factor of $10$.
Table~\ref{t1} shows a comparison between PS191 and several examples of
recent neutrino experiments.
Among several options of accelerator experiments, in this paper we focus on 
the T2K experiment as a typical example.
%%%%%%%%%%%%%%%%%%%%%%%%%%%%%%%%%%%%%%%%%%%%%%%%%%%%%%%%%%%%%%%%%%%%%%%%%%%%%%%
\begin{table}
\begin{center}
\begin{tabular}{cccccc}\hline
 & PS191~\cite{PS191} & T2K~\cite{Abe:2011ks} & MINOS~\cite{MINOS} & 
MiniBooNE~\cite{Mini} & SciBooNE~\cite{Sci}\\\hline
POT & $0.86 \times 10^{19}$ & $10^{21}$ & $10^{21}$ & $10^{21}$ & $10^{21}$\\
$({\rm Distance})^{-2}$ & $(128\,{\rm m})^{-2}$ & $(280\,{\rm m})^{-2}$ &
$(1\,{\rm km})^{-2}$ & $(541\,{\rm m})^{-2}$ & $(100\,{\rm m})^{-2}$ \\
Volume & $216\,{\rm m}^3$ & $88\,{\rm m}^3$ &$303\,{\rm m}^3$ &$524\,{\rm m}^3$ 
&$15.3\,{\rm m}^3$ \\
Events & $1$ & $9.9$ & $2.7$ & $15.8$ & $13.5$ \\\hline
\end{tabular}
\caption{A comparison between PS191 and recent accelerator experiments.
The item ``Distance'' means the distance between the beam target and the detector
for each experiment.
The item ``Events'' shows POT$\times$(Distance)${}^{-2}$$\times$Volume in units 
of PS191. The POTs for the oscillation experiments are assumed to achieve
$10^{21}$.}
\label{t1}
\end{center}
\end{table}
%%%%%%%%%%%%%%%%%%%%%%%%%%%%%%%%%%%%%%%%%%%%%%%%%%%%%%%%%%%%%%%%%%%%%%%%%%%%%%%

The study follows two main steps;~the flux estimation and the event number calculation 
for various signal decays.
First, in the fulx estimation, we use a semi-analytical method with a help of 
the active neutrino flux $\phi_\nu$ simulated by the T2K collaboration~\cite{flux,D1,
Abe:2012av},
making a reasonable simplification for the geometry of the decay tunnel and the
detector.
More precise fluxes of the heavy neutrino might be obtained by Monte Carlo methods.
In a Monte Carlo calculation it is possible to take into account the details of the
geometry and the spectrum of the parent mesons.
The analytical technique is nevertheless useful, since it allows one to understand 
the essential physics which determines the behavior of the heavy neutrino spectrum,
in particular its mass dependence.
We emphasize that the phase-space effect in kaon decay is important 
and the heavy neutrino flux 
$\phi_N$ can be significantly deviated from the naive expectation $\phi_N \simeq 
|\Theta |^2 \phi_\nu$, where $\Theta$ is the active-heavy mixing parameter 
of interest.
As the heavy neutrino mass $M_N$ approaches to the production threshold, 
the heavy neutrinos tend to be distributed to lower energies with a narrow spread,
so that the proportionality 
$\phi_N \simeq |\Theta |^2 \phi_\nu$ is broken.
This phase-space effect leads to larger event numbers than the naive expectation
and has a significant impact for the estimation of the sensitivity.

Second, in the event number calculation, we take into account various decay modes
of the heavy neutrino $N$.
The two-body modes $N \to e^\mp \pi^\pm$ and $N \to \mu^\mp \pi^\pm$ are the
most promising channels due to their large branching ratios.
For these modes, the invariant mass distribution for the lepton and the pion momenta 
has a peak at the heavy neutrino mass.
The three-body modes $N \to e^- e^+ \nu$, $N \to \mu^\mp e^\pm \nu$ and
$N \to \mu^- \mu^+ \nu$ are also interesting.
While the first mode may suffer higher background by $\pi^0$ decay, the latter
two modes have smaller backgrounds and also serve as promising channels 
for discovering the heavy neutrino.
By assuming non-observation of these exotic events in the T2K near detector
at $10^{21}\,{\rm POT}$, one finds the upper-bound for the mixing parameter
of the electron type better than that of PS191.
This means that T2K may have a good chance to discover the heavy neutrino in
near future.

The layout of this paper is as follows.
In Section 2, we introduce the heavy neutrino and  briefly  review its essential 
properties.
In Section 3, the flux of the heavy neutrino at the near detector is discussed.
In Section 4, the decays of the heavy neutrino at the detector and their event numbers
are studied.
Section 5 is devoted to conclusions.

%%%%%%%%%%%%%%%%%%%%%%%%%%%%%%%%%%%%%%%%%%%%%%%%%%%%%%%%%%%%%%%%%%%%%%
\section{Properties of the heavy neutrino}
\label{decay}
%%%%%%%%%%%%%%%%%%%%%%%%%%%%%%%%%%%%%%%%%%%%%%%%%%%%%%%%%%%%%%%%%%%%%%
We consider a heavy (sterile) neutrino $N$ in the mass range 
$1\,{\rm MeV} \lesssim M_N \lesssim 500\,{\rm MeV}$.
The flavor eigenstates of the left-handed neutrinos $\nu_\alpha$ ($\alpha = e,\mu,\tau$) 
are given by the linear combination of the mass eigenstates as
\begin{eqnarray}
\nu_\alpha \,=\, U_{\alpha i} \,\nu_i + \Theta_{\alpha}N,
\end{eqnarray}
where $U_{\alpha i}$ is the Pontecorvo-Maki-Nakagawa-Sakata (PMNS) matrix,
$\nu_i$ ($i = 1,2,3$) are the mass eigenstates of the active neutrinos.
The parameter $\Theta_{\alpha}$ is the mixing between (light) active and sterile 
neutrinos, through which $N$ interacts with the weak gauge bosons.
The extension to the multi-generation case is trivially done
by replacing $\Theta_{\alpha}N$ with $\sum_I \Theta_{\alpha I}N_I$.
In this paper, we assume $N$ is Dirac particle unless otherwise stated.
This is simply because we would like to make a comparison to PS191 in which 
the same assumption is made. 
If $N$ is Majorana particle, the decay width is doubled since $N$ decays to 
charge-conjugated states also.

Let us overview the way to find the heavy neutrino $N$ in accelerator
experiments, especially in the T2K experiment.
In the mass range of $140\,{\rm MeV} \lesssim M_N \lesssim 500\,{\rm MeV}$, 
the heavy neutrino $N$ is produced by $K$ decay.
The main production modes are
\begin{eqnarray}
K^+ \to \mu^+ N,\quad\quad K^+ \to e^+ N
\label{KtoN}
\end{eqnarray}
for $M_N < 388\,{\rm MeV}$ and $M_N < 493\,{\rm MeV}$, respectively.
The decay width $K^+ \to \mu^+ N$ ($K^+ \to e^+ N$) is proportional to 
$|\Theta_\mu |^2$ ($|\Theta_e |^2$), and $\Theta_\tau$ is irrelevant for the
production\footnote{The heavier mesons such as $D$ can decay producing tau
so that $\Theta_\tau$ is involved in the production. In this work we neglect the
contribution from such heavier mesons.}.
We describe further details of the decay modes~(\ref{KtoN}) in Section 3 and 
Appendix A.
Since the magnetic horn focuses positively-charged mesons, the contributions 
from $K^-$ are small~\cite{D1} and in what follows we neglect the $K^-$ contributions.

The heavy neutrinos $N$ produced by the meson decays escape from the decay volume
and some of them are injected into the near detector ND280 in the T2K experiment 
and decay to leave signals.
Depending on the mass, $N$ decays to lighter particles in various decay modes; 
\begin{gather}
N \to \gamma \nu,\quad N \to 3\nu,\quad
N \to e^- e^+ \nu, \quad N \to \mu^\mp e^\pm \nu,\quad\nonumber\\
N \to \nu \pi^0,\quad\
N \to e^- \pi^+,\quad N \to \mu^- \mu^+ \nu,\quad N \to \mu^- \pi^+.
\nonumber
\end{gather}
With the ND280 detector capable of identifying $e^\pm, \mu^\pm$ and $\pi^\pm$, 
some of the above channels can be detected as signal events.

Due to large branching ratios, the two-body decay modes $N \to \nu \pi^0$,
$N \to e^- \pi^+$ and $N \to \mu^- \pi^+$ would be the most frequent events.
However, $N \to \nu \pi^0$ does not seems to be as promising as the other two modes since
$\pi^0$ are copiously produced by the ordinary neutrino interactions, which lead to
large backgrounds.
On the other hand, $N \to e^- \pi^+$ and $N \to \mu^- \pi^+$ leave
two charged-particle tracks with monochromatic energies in the rest frame of $N$.
The invariant mass distribution of the two charged-particle momenta sharply peaks 
at the heavy neutrino mass.
Such a peak signal is definitely better than a slight excess of $\pi^0$ events 
over huge background.
We thus proceed without any further analysis of $N \to \nu \pi^0$, but
study $N \to e^- \pi^+$ and $N \to \mu^- \pi^+$ in more detail in Section 4.
The decay rates of $N \to e^- \pi^+$ and $N \to \mu^- \pi^+$ are proportional
to $|\Theta_e|^2$ and $|\Theta_\mu|^2$, respectively.
The radiative decay $N \to \gamma \nu$ is induced at one loop and negligible 
in this work.

In spite of small branching ratios, the three-body decay modes $N \to e^- e^+ \nu$,
$N \to \mu^\mp e^\pm \nu$ and $N \to \mu^- \mu^+ \nu$ are also interesting to study.
In particular, $N \to \mu^\mp e^\pm \nu$ and $N \to \mu^- \mu^+ \nu$ have smaller 
backgrounds than $N \to e^- e^+ \nu$ and a few these events may lead to the 
discovery of the heavy neutrino.
Notice that $N \to e^- e^+ \nu$ and $N \to \mu^- \mu^+ \nu$ are conducted not 
only by the charged-current (CC) interaction but also by the neutral current (NC)
interaction, so that these decay rates depend on all types of the mixing parameters
$\Theta_{e,\mu,\tau}$.
On the other hand, $N \to \mu^\mp e^\pm \nu$ are mediated only by the CC interactions
and the decay rate depends on $\Theta_{e,\mu}$.

\begin{figure}[t]
\begin{center}
\scalebox{0.42}{\includegraphics{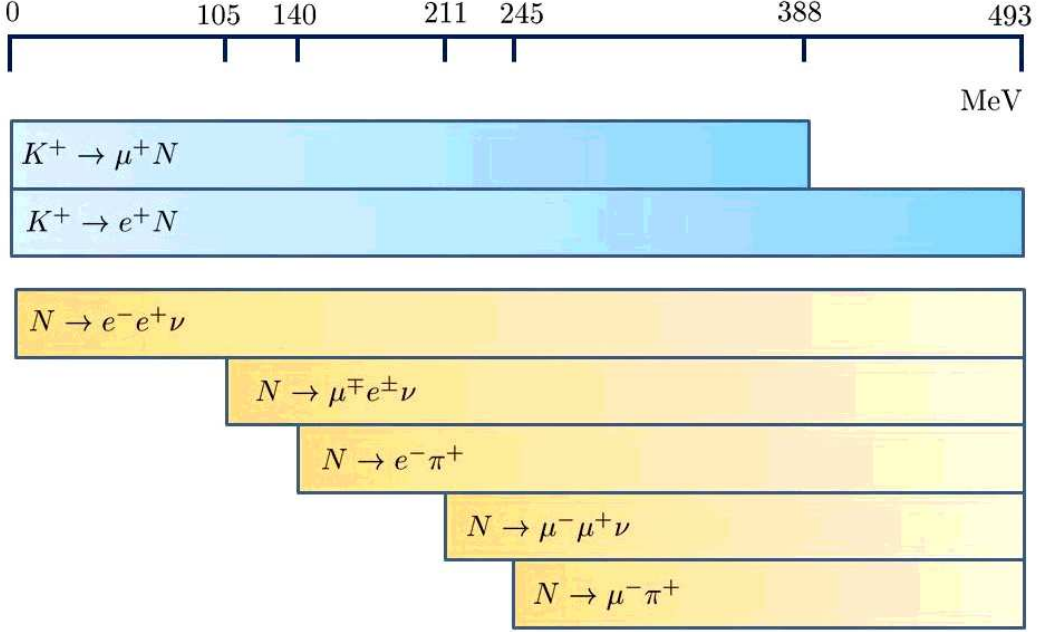}}
\end{center}
\caption{Summary of the production and the detection processes of the heavy neutrino $N$.
The decay mode $N \to \nu \pi^0$ is abbreviated here because it is not as
promising detection channel as the other ones.}
\label{PandD}
\end{figure}

Fig.~\ref{PandD} summarizes the production and decay processes of $N$ to
be studied in this work.
We do not present the formulas for the decay rates here.
A complete list of the decay modes and the decay rates is found 
in Ref.~\cite{Gorbunov:2007ak}.

%%%%%%%%%%%%%%%%%%%%%%%%%%%%%%%%%%%%%%%%%%%%%%%%%%%%%%%%%%%%
%%%%%%%%%%%%%%%%%%%%%%%%%%%%%%%%%%%%%%%%%%%%%%%%%%%%%%%%%%%%

%%%%%%%%%%%%%%%%%%%%%%%%%%%%%%%%%%%%%%%%%%%%%%%%%%%%%%%%%%%%%%%%%%%%%%
\section{Heavy neutrino flux at the near detector}
\label{flux}
\begin{figure}[t]
\begin{center}
\scalebox{0.4}{\includegraphics{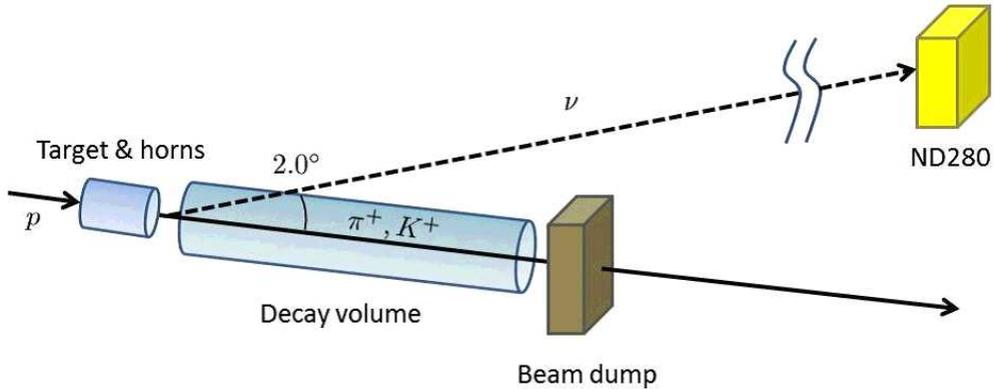}}
\end{center}
\caption{Schematic of the secondary beam line and the near detector ND280.}
\label{layout}
\end{figure}

%%%%%%%%%%%%%%%%%%%%%%%%%%%%%%%%%%%%%%%%%%%%%%%%%%%%%%%%%%%%%%%%%%%%%%
In the T2K experiment, pions and kaons are produced by the interaction of
$31\,{\rm GeV}$ protons with the graphite target.
The produced mesons are focused by the magnetic horns and enter the decay volume
of $96\,{\rm m}$ long filled with helium gas.
The parent mesons decay in flight inside the decay volume.
The off-axis detector ND280 is located $280\,{\rm m}$ from the target station.
The off-axis angle to ND280 from the target position is $2.04^\circ$. 
%$2.04^\circ$. 
The layout of the secondary beam line and the near detector is sketched 
in Fig.~\ref{layout}.
Details of the experiment setup are found in Ref.~\cite{Abe:2012av}.
%%%%%%%%%%%%%%%%%%%%%%%%%%%%%%%%%%%%%%%%%%%%%%%%%%%%%%%%%%%%
%%%%%%%%%%%%%%%%%%%%%%%%%%%%%%%%%%%%%%%%%%%%%%%%%%%%%%%%%%%% 

Calculation of the neutrino flux at the near detector is a complicated task.
The T2K collaboration has simulated the fluxes of the active neutrinos by the 
Monte-Carlo method.
In this work, we do not follow their approach, but estimate the heavy neutrino
flux by a semi-analytical method similar to Ref.~\cite{Lipari}.
We try to reconstruct a reasonable flux of parent particles from the active
neutrino flux of Ref.~\cite{D1}, and then evaluate the heavy neutrino flux from the 
reconstructed parent flux. 
In this paper, we focus on $K^+$ meson as the parent of the heavy neutrinos since 
$K^+$ decay covers a wider range of the heavy neutrino mass than $\pi^+$ decay.
The following discussion is, however, easily extended to the $\pi^+$ case.

%%%%%%%%%%%%%%%%%%%%%%%%%%%%%%%%%%%%%%%%%%%%%%%%%%%%%%%%%%%%%%%%%%%%%%
\subsection{Modeling the parent flux}
\label{recon}
%%%%%%%%%%%%%%%%%%%%%%%%%%%%%%%%%%%%%%%%%%%%%%%%%%%%%%%%%%%%%%%%%%%%%%

For ND280, the neutrino source is a line-like object rather than a point-like one.
Let $\phi(p_K,l)$ denotes the $K^+$ spectrum along the decay volume, where
$p_K$ is the magnitude of the $K^+$ momentum and $l$ is the flight length of $K^+$.
We set $l=0$ to be the upstream end of the decay volume.
In the decay volume filled with helium gas, the decay length of $K^+$ is much 
shorter than the interaction length.
One thus finds
\begin{eqnarray}
\phi_K(p_K,l) \,=\, \phi_K(p_K)e^{-\frac{l}{\Lambda_K}},
\end{eqnarray}
where $\phi_K(p_K)$ is the spectrum at $l=0$ and $\Lambda_K$ is the total decay length
$\Lambda_K = 3.7 (p_K/m_K)\,{\rm m}$ with the kaon mass $m_K = 493\,{\rm MeV}$.

If the parent spectrum $\phi_K(p_K)$ is known, one can calculate the daughter $\nu_\mu$ 
flux from $K^+ \to \mu^+ \nu_\mu$ decay.
The source term of $\nu_\mu$ is given by
\begin{eqnarray}
S_\nu(E_\nu, \theta, \phi, l) &=& 
\int_0^\infty \!\!dp_K \,\frac{\phi_K(p_K,l)}{\beta (1/\Gamma) \gamma} \,\,
\frac{1}{\Gamma}\frac{d^3 \Gamma}{dE_\nu d\cos\theta d\phi}\nonumber\\
&=& \int_0^\infty \!\!dp_K \,\phi_K(p_K,l)\left( \frac{m_K}{p_K} \right)
\frac{d^3 \Gamma}{dE_\nu d\cos\theta d\phi},
\label{Snu}
\end{eqnarray}
where $E_\nu$ is $\nu_\mu$ energy, $\theta$ and $\phi$ are the polar and
the azimuth angles of the emitted $\nu_\mu$ relative to the $K^+$ momentum directions, 
and $\Gamma$ is the $K^+ \to \mu^+ \nu_\mu$ decay width.
Giving the source term, the $\nu_\mu$ flux $\phi_{\nu_\mu}(E_\nu)$ at ND280
is obtained by integrating the source term along $l$ and the angles $\theta$ and $\phi$
covered by ND280.
It reads
\begin{eqnarray}
\phi_{\nu_\mu}(E_\nu) \,=\, \int_0^{l_f}\!\!\! dl \int_{-1}^{1} \!\!d\!\cos\theta 
\int_{0}^{2\pi} \!\!\!d\phi \,
\frac{1}{A}\,S_\nu(E_\nu, \theta, \phi, l)\,P(\theta, \phi),
\label{phi1}
\end{eqnarray}
where $l_f = 96\,{\rm m}$, $A$ is the effective area of ND280, $P(\theta,\phi)$ 
is the ``projection'' function which turns unity if $\nu_\mu$ enters ND280 
and otherwise zero.

The parent kaons carry two more degrees of freedom which are not
explicitly mentioned above; the polar angle relative
to the beam axis (the central axis of the decay volume) and the radial coordinate
from the beam axis, both of them are defined at $l=0$.
The location of ND280 relative to each $K^+$ momentum depends on these degrees of 
freedom and the function $P(\theta,\phi)$ is in fact highly complicated.
However, the situation is greatly simplified if kaon momenta are assumed to be
parallel to the beam axis. 
Information about the polar angle and the radial coordinate is then to be represented 
by an effective off-axis angle $\theta_0$, which is an ``virtual'' off-axis angle 
to ND280 from the upstream end of the decay volume. 
This angle $\theta_0$ represents the average off-axis angle to ND280 from each
$K^+$ momentum, which angle varies kaon by kaon carrying deferent polar angles
and radial coordinates.
Furthermore, we neglect $\theta$ dependence of the effective area $A$.
Adopting these simple modeling of geometry of the decay volume and the detector,
we use
\begin{eqnarray}
\phi_{\nu_\mu}(E_\nu) \,=\, \frac{\Delta \phi}{A}
\int_0^{l_f}\!\!\! dl \int_{-1}^{1} \!\!d\!\cos\theta 
\,S_\nu(E_\nu, \theta, l)\,P'(\theta,\theta_0),
\label{phi2}
\end{eqnarray}
as a formula to relate the daughter $\nu_\mu$ flux with the parent $K^+$ spectrum.
Here $\Delta \phi$ is $\phi$ interval defined by the detector width, 
$P'(\theta,\theta_0)$ denotes the projection function determined by
the detector hight and $K^+$ decay point $l$.
Since the explicite expression of Eq.~(\ref{phi2}) is rather lengthy, we put 
details on Eq.~(\ref{phi2}) in Appendix~\ref{fluxculc} .
%%%%%%%%%%%%%%%%%%%%%%%%%%%%%%%%%%%%%%%%%%%%%%%%%%%%%%%%%%%%
\begin{figure}[t]
    \begin{center}
\scalebox{0.6}{\includegraphics{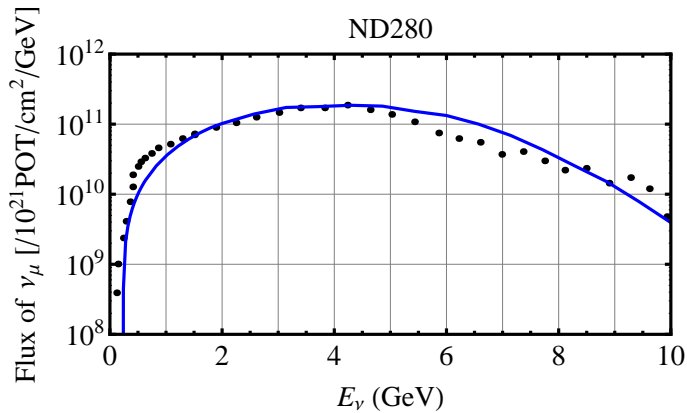}}
\end{center}
\caption{A comparison between Eq.~(\ref{phi2}) and the $\nu_\mu$ flux simulated 
in Ref.~\cite{D1}. 
Dots shows the result of the simulation in Ref.~\cite{D1} and
solid curve is Eq.~(\ref{phi2}) with the parameters $\theta_0 = 1.48^\circ$,
$a_0 = 4.8\times 10^{19}\,{\rm mb^{-1}}$ and $p_0 = 2.1\,{\rm GeV}$.}
\label{fND280}
\end{figure}
%%%%%%%%%%%%%%%%%%%%%%%%%%%%%%%%%%%%%%%%%%%%%%%%%%%%%%%%%%%%

Having Eq.~(\ref{phi2}), our strategy is fitting $\phi_{\nu_\mu}(E_\nu)$ calculated
in Ref.~\cite{D1} by adjusting the parameters included in the right-handed side 
of Eq.~(\ref{phi2}), with a kaon spectrum which is physically well-motivated. 
The $K^+$ spectrum $d\sigma/dp_K$ in the proton collision with a graphite target 
is measured by NA61/SHINE Collaboration~\cite{Abgrall:2011ae}, which is customized
to improve the flux calculation in T2K.
It seems reasonable to expect that the shape of the kaon spectrum is not far
from this measured spectrum.
However, the effects such as the secondary protons~\cite{D1} and the magnetic horns
must deform $d\sigma/dp_K$ to some extent.
To take into account this deformation, we allow a shift of the peak of $d\sigma/dp_K$.
In summary,  in order to model the kaon spectrum $\phi_K(p_K)$, we introduce two free 
parameter $a_0$ and $p_0$, the overall scale factor and the shift of the peak, 
respectively. 
See Appendix~\ref{fluxculc} for more details.

With the above ansatz for the kaon spectrum, we have three parameters in the 
right-hand side of Eq.~(\ref{phi2}); the effective off-axis angle $\theta_0$,
the scale factor $a_0$ and the shift of the peak $p_0$.
Fig.~\ref{fND280} shows the fit by Eq.~(\ref{phi2}).
The dots show the result of the simulation in Ref.~\cite{D1}, while the solid curve 
shows Eq.~(\ref{phi2}) with the parameter set of $\theta_0 =  1.48^\circ$,
$a_0 = 4.8\times 10^{19}\,{\rm mb^{-1}}$ and $p_0 = 2.1\,{\rm GeV}$.
It is seen that Eq.~(\ref{phi2}) is well tracing the global behavior
of the simulated result of Ref.~\cite{D1}.

Notice that the effective off-axis angle is taken as $\theta_0 = 1.48^\circ$, 
which is smaller than the actual one $2.04^\circ$.
If this angle is taken as $\theta_0 = 2.04^\circ$, the flux starts to fall off at
$E_\nu \sim 6\,{\rm GeV}$ and does not fit the data points above $6\,{\rm GeV}$.
This is simply due to the two-body kinematics.
In $K^+ \to \mu^+ \nu_\mu$ decay, the maximum energy of $\nu_\mu$ is
\begin{eqnarray}
E^{\rm max}_\nu = \frac{m_K^2 - m_\mu^2}{2m_K \sin\theta}
\end{eqnarray}
for $\theta < 90^\circ$.
The larger the off-axis, the smaller the maximum neutrino energy. 
The abundance of $\nu_\mu$ above $6\,{\rm GeV}$ means that certain fraction of
kaons must have momentum directing to the near detector so that $\nu_\mu$ with
smaller $\theta$ can contribute to the flux.
This is of course expected since the magnetic horns do not make the kaon
momenta perfectly parallel to the beam axis.
The smaller angle $\theta_0 = 1.48^\circ$ effectively takes this effect into account.

%%%%%%%%%%%%%%%%%%%%%%%%%%%%%%%%%%%%%%%%%%%%%%%%%%%%%%%%%%%%%%%%%%%%%%
\subsection{The heavy neutrino flux}
\label{Hflux}
%%%%%%%%%%%%%%%%%%%%%%%%%%%%%%%%%%%%%%%%%%%%%%%%%%%%%%%%%%%%%%%%%%%%%%
%%%%%%%%%%%%%%%%%%%%%%%%%%%%%%%%%%%%%%%%%%%%%%%%%%%%%%%%%%%%
\begin{figure}[t]
    \begin{center}
%\scalebox{0.04}{\includegraphics{Nfluxmu_eL_rev2.eps}}
\scalebox{0.6}{\includegraphics{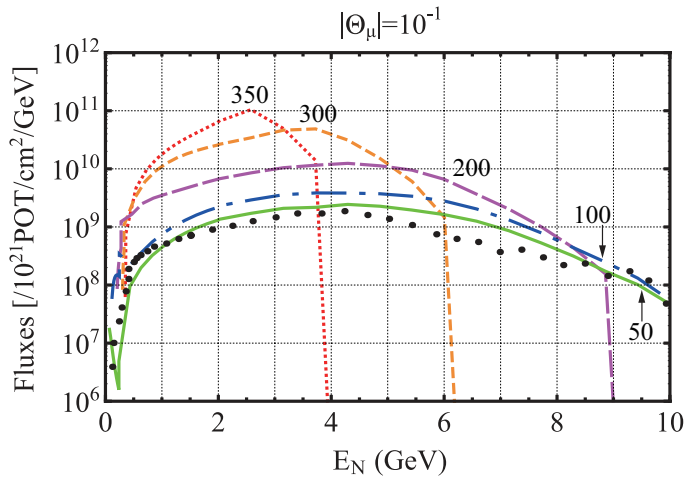}}
\end{center}
\caption{Fluxes of the heavy neutrino $\phi_N(E_N)$ from the $K^+ \to \mu^+ N$ mode
for several sample values
of $M_N$. Black dotted marks show $|\Theta_\mu |^2 \phi_{\nu_\mu}$.
The masses are taken as $M_N = 350\,{\rm MeV}$ (red, dotted), 
$300\,{\rm MeV}$ (orange, dashed), $200\,{\rm MeV}$ (magenda, long-dashed),
$100\,{\rm MeV}$ (blue, dashed-dot), $50\,{\rm MeV}$ (green, solid).}
\label{Nflux1}
\end{figure}
%%%%%%%%%%%%%%%%%%%%%%%%%%%%%%%%%%%%%%%%%%%%%%%%%%%%%%%%%%%%

With the kaon spectrum $\phi_K(p_K)$ discussed above, 
let us estimate the heavy neutrino flux $\phi_N(E_N)$.
The calculation goes as before.
The source term is given by
\begin{eqnarray}
S_N(E_N, \theta, \phi, l) \,=\, 
\int_0^\infty \!\!dp_K \,\phi_K(p_K,l)\left( \frac{m_K}{p_K} \right)
\sum_{i=1}^2
\frac{d^3 \Gamma_i}{dE_N d\!\cos\theta d\phi},
\label{SN}
\end{eqnarray}
where $\Gamma_1$ and $\Gamma_2$ are the decay width for $K^+ \to \mu^+ N$ and
$K^+ \to e^+ N$, respectively.
Provided that the decay length of the heavy neutrino is much larger than the
distance between the $K^+$ decay point and the detector, 
the heavy neutrino flux $\phi_N(E_N)$ is given by just replacing $S_\nu$ with
$S_N$ in Eq.~(\ref{phi2}).
We will see in Section~\ref{Eventrate} that this is the case for
the energy and the parameters of the current interests.
Details on the differential decay rates $d\Gamma_i$ are shown 
in Appendix~\ref{fluxculc}.

%%%%%%%%%%%%%%%%%%%%%%%%%%%%%%%%%%%%%%%%%%%%%%%%%%%%%%%%%%%%
\begin{figure}[t]
    \begin{center}
%\scalebox{0.04}{\includegraphics{Nfluxe_eL_rev2.eps}}
\scalebox{0.6}{\includegraphics{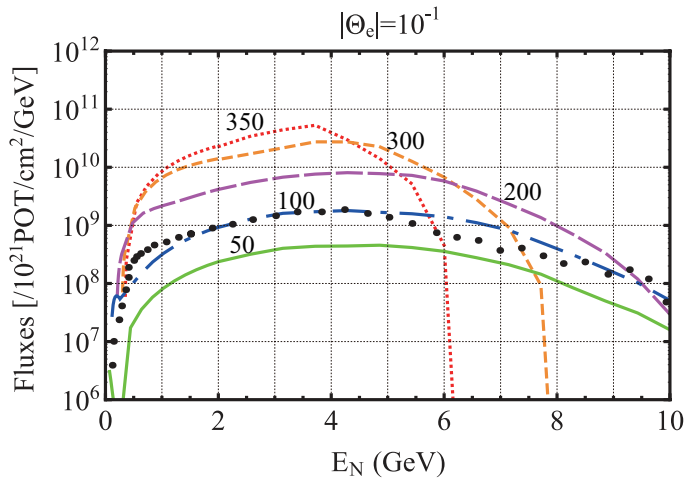}}
\end{center}
\caption{Same as Fig.~\ref{Nflux1} but for the $K^+ \to e^+ N$ mode.}
\label{Nflux2}
\end{figure}
%%%%%%%%%%%%%%%%%%%%%%%%%%%%%%%%%%%%%%%%%%%%%%%%%%%%%%%%%%%%

Fig.~\ref{Nflux1} and~\ref{Nflux2} present the heavy neutrino fluxes $\phi_N(E_N)$
for several fixed values of $M_N$.
Fig.~\ref{Nflux1} and~\ref{Nflux2} show the cases where 
$(|\Theta_e| ,|\Theta_\mu|) = (0,10^{-1}) $ and $(10^{-1},0)$,
respectively.
% The masses are sampled as
% $M_N = 350\,{\rm MeV}\,(\rm purple)$, 
% $300\,{\rm MeV}\,(\rm blue)$, $200\,{\rm MeV}$ $(\rm cyan)$,
% $100\,{\rm MeV}\,(\rm green)$, $50\,(\rm orange)\,{\rm MeV}$.
In the both cases of $K^+ \to \mu^+ N$ and $K^+ \to e^+ N$, 
the spectral shapes are similar to $\phi_{\nu_\mu}$ for $M_N \lesssim 100\,{\rm MeV}$, 
whereas they are significantly deviated from $\phi_{\nu_\mu}$ 
for $M_N \gtrsim 100\,{\rm MeV}$.
The most remarkable feature is the enhancement of the flux at lower energies for 
larger $M_N$.
In Fig.~\ref{Nflux1}, for example, the heavy neutrinos of $M_N = 350\,{\rm MeV}$ gather 
around $2-3\,{\rm GeV}$ and the peak intensity reaches $10^{11}\,{\rm cm^{-2} GeV^{-1}}$.
This is nearly two orders of magnitude larger than the naive expectation 
$|\Theta_\mu|^2 \phi_{\nu_\mu}$.

There are two reasons for the enhancement.
One is the spin conservation in the $K^+$ rest frame. 
The matrix elements of the decay processes $K^+ \to \mu^+ N$ and $K^+ \to e^+ N$
scale with $M_N^2$ when $M_N$ is much larger than the mass of the charged lepton in
each final state.
This accounts for the larger fluxes than $|\Theta_{e,\mu}|^2 \phi_{\nu_\mu}$ 
for $M_N > m_\mu$.
This also accounts for the smaller flux than $|\Theta_e|^2 \phi_{\nu_\mu}$
for $M_N = 50\,{\rm MeV}$ in Fig.~\ref{Nflux2}.

The another reason is slower motions of the heavy neutrinos
at the rest frame of the parent particle. 
The smaller the daughter velocities, the easier to boost them into the forward
directions.
In the current setup, the detector may well be regarded as the object placed in
the forward direction of the $K^+$ momenta. 
In the case of $K^+ \to \mu^+ \nu_\mu$, neutrinos can be emitted not only to
the forward directions but also to the backward directions since the neutrino masses 
are very small and 
the neutrino velocities at the rest frame are larger than the parent typical velocity 
at the laboratory frame.
On the other hand, in the decay $K^+ \to \mu^+ N$, the heavy neutrinos tend to
be emitted to the forward directions.
In the rest frame of $K^+$, the gamma factor of $N$ at the $K^+$ rest frame is given by
\begin{eqnarray}
\gamma_N = \frac{m_K^2 - m_\mu^2 + M_N^2}{2m_K M_N}.
\label{eN}
\end{eqnarray} 
This goes to unity as $M_N$ approaches to the threshold value $M_N = m_K - m_\mu
= 388\,{\rm MeV}$.
On the other hand, most of the kaons carry the momentum around $1-4\,{\rm GeV}$,
so that the gamma factor of $K^+$ is typically given by 
$\gamma_K = 2-8$. 
Then Eq.~(\ref{eN}) tells us that $\gamma_K > \gamma_N$ for $M_N \gtrsim 120\,{\rm MeV}$.
Hence for $M_N \gtrsim 120\,{\rm MeV}$, kaon's velocities overcome $N$'s velocities
and $N$ are focused into the forward directions.
This agrees with the flux behavior seen in Fig.~\ref{Nflux1} and~\ref{Nflux2}.

%%%%%%%%%%%%%%%%%%%%%%%%%%%%%%%%%%%%%%%%%%%%%%%%%%%%%%%%%%%%%%%%%%%%%%
\section{Event rates and expected sensitivity}
\label{Eventrate}
%%%%%%%%%%%%%%%%%%%%%%%%%%%%%%%%%%%%%%%%%%%%%%%%%%%%%%%%%%%%%%%%%%%%%%
As we have seen in Section~\ref{decay}, a fraction of the heavy neutrinos passing 
through the detector decay inside the detector volume and leave signals via
various decay modes.
In this section, we calculate the number of signal events in ND280 and
estimate the potential sensitivity of ND280. 
We argue that the sensitivity of ND280 is comparable to that of the PS191 experiment.

%%%%%%%%%%%%%%%%%%%%%%%%%%%%%%%%%%%%%%%%%%%%%%%%%%%%%%%%%%%%%%%%%%%%%%
\subsection{Number of the signal events}
%%%%%%%%%%%%%%%%%%%%%%%%%%%%%%%%%%%%%%%%%%%%%%%%%%%%%%%%%%%%%%%%%%%%%%
The total number of events is given by the difference between the number of
the heavy neutrinos at the up and down stream  end of the detector.
For a particular decay channel, the number of events is given by
\begin{eqnarray}
{\rm Events} = A \int_{M_N}^\infty \!\! dE_N \,\,\frac{1}{\lambda}\int_{x_0}^{x_1}\!\!\! 
dx \,\,
\phi_N(E_N,x),
\label{Eventmaster}
\end{eqnarray}
where $\lambda$ is the (partial) decay length for the signal decay mode of interest,
$A$ is the cross-sectional area of the detector, $x$ is the flight distance 
of the heavy neutrinos and $(x_0, x_1)$ means the detector segment.
The number of the heavy neutrinos decreases by the decays.
With the total decay length $\Lambda_N$, the $x$ dependence of $\phi_N(E_N, x)$
is determined as
\begin{eqnarray}
\phi_N(E_N, x) \,=\, \phi_N(E_N)e^{-\frac{x}{\Lambda_N}},
\label{phitot}
\end{eqnarray}
where $\phi_N(E_N)$ is the heavy neutrino flux discussed in Section~\ref{Hflux}.

Eq.~(\ref{Eventmaster}) is further simplified if the total decay length is
much larger than the flight distance of the heavy neutrino.
Provided that $\Lambda_N \gg x_0, x_1 - x_0$, the number of events reads
\begin{eqnarray}
{\rm Events} \,\simeq\,  \int_{M_N}^\infty \!\! dE_N \,\,
\phi_N(E_N)\,\frac{V}{\lambda},
\label{Eventmaster4}
\end{eqnarray}
where $V = A(x_1 - x_0)$ is the detector volume. 
In the T2K setup, $x_0 \approx 300\,{\rm m}$ and $x_1 - x_0 \approx 5\,{\rm m}$.
Thus the condition $\Lambda_N \gg x_1 - x_0$ holds if $\Lambda_N \gg x_0 \approx 
300\,{\rm m}$.

It turns out that the condition $\Lambda_N \gg 300\,{\rm m}$ holds in
the most parameter and energy region of interest.
%%%%%%%%%%%%%%%%%%%%%%%%%%%%%%%%%%%%%%%%%%%%%%%%%%%%%%%%%%%%%%%%%%%%%%
\begin{figure}[t]
  \centerline{
  \includegraphics[scale=0.28]{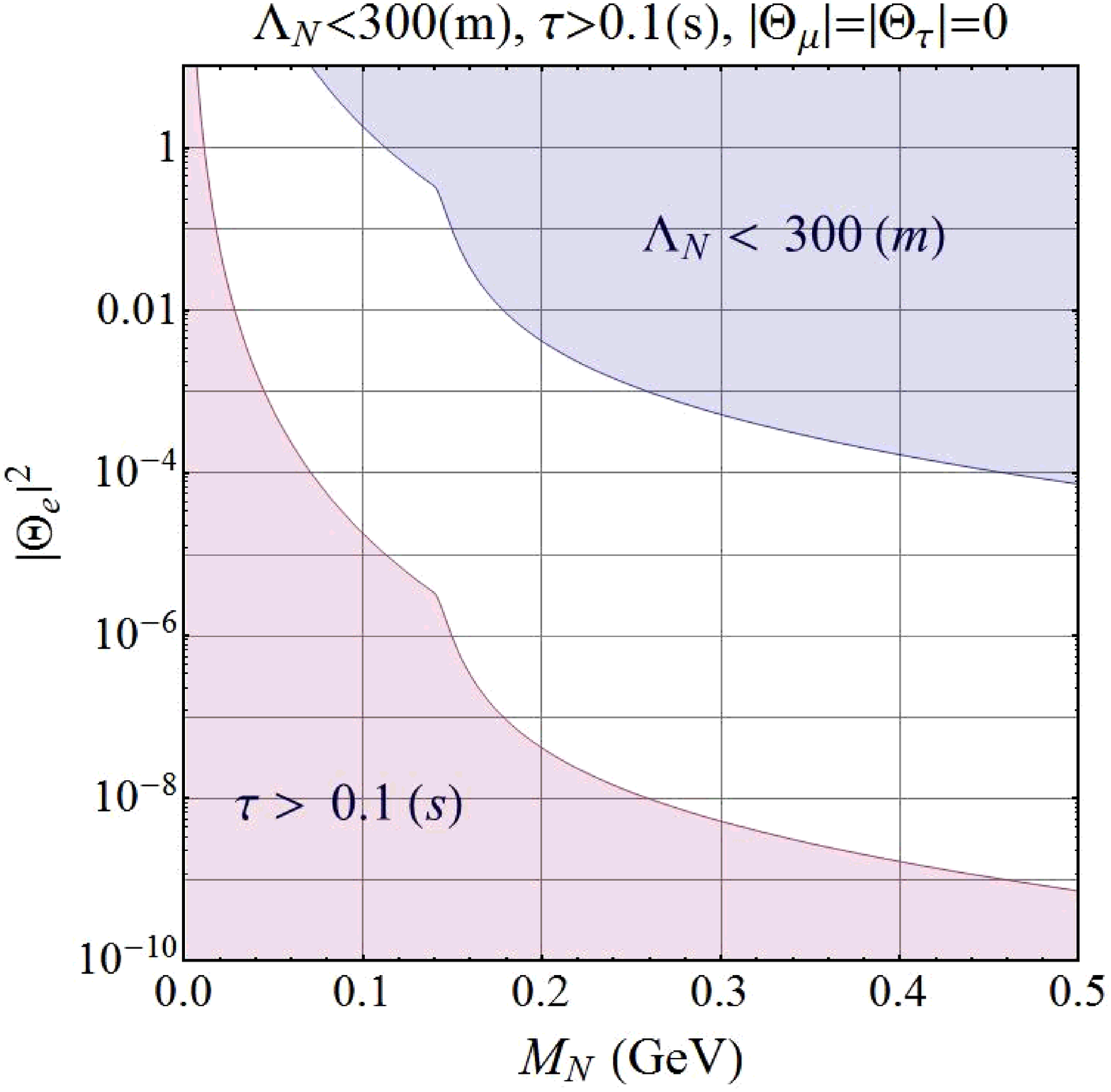}%
  \includegraphics[scale=0.28]{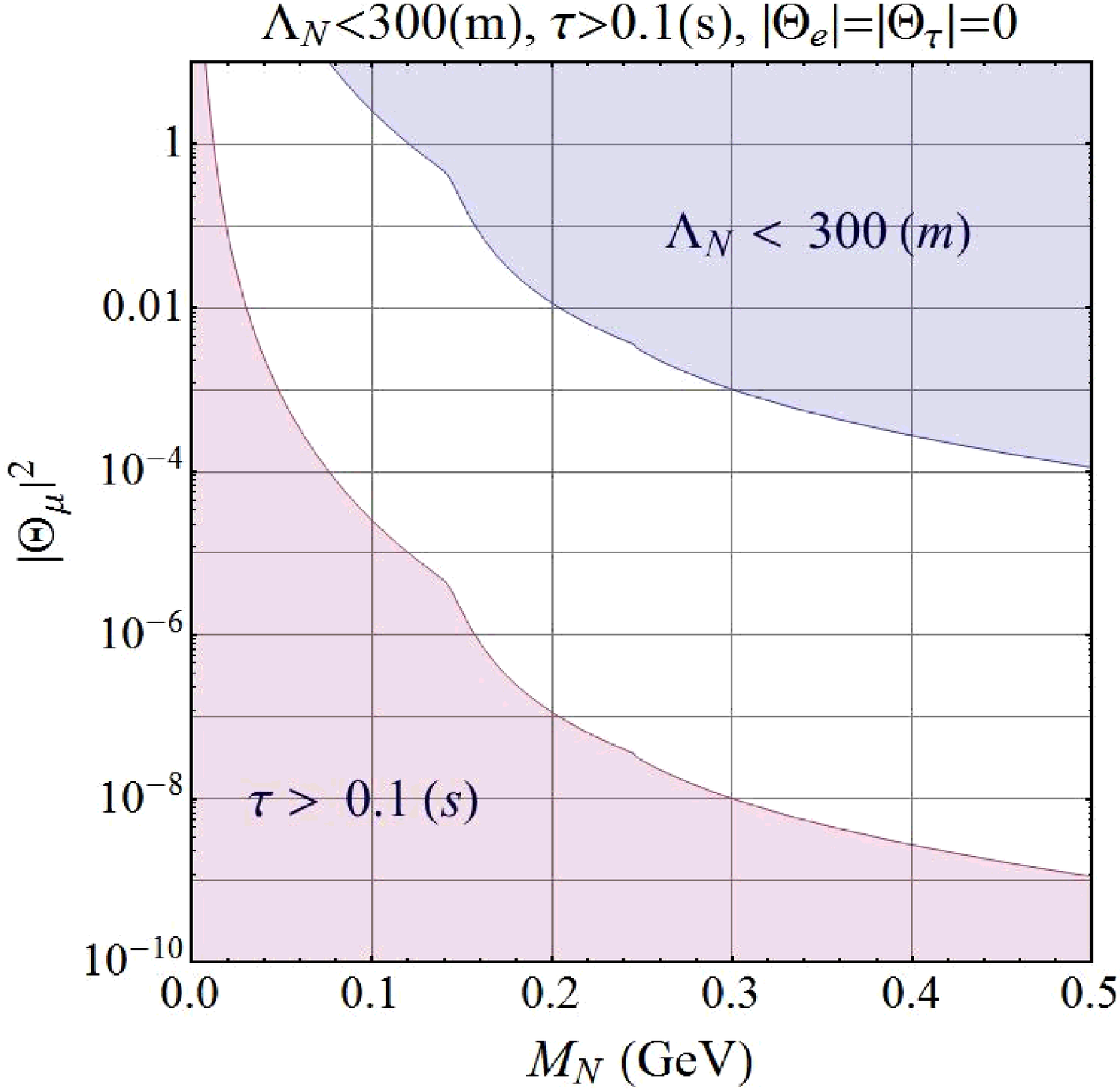}%
  }%
\caption{Parameter region that satisfy $\Lambda_N < 300\,{\rm m}$ for 
$\gamma_N = \sqrt{2}$ (The filled region labeled by $\Lambda_N < 300\,{\rm m}$).
The filled region labeled by $\tau > 0.1 (\rm s)$ shows the region where
heavy neutrino decay may spoil BBN.}
  \label{tot}
\end{figure}
%%%%%%%%%%%%%%%%%%%%%%%%%%%%%%%%%%%%%%%%%%%%%%%%%%%%%%%%%%%%%%%%%%%%%% 
Fig.~\ref{tot} highlights a ``strong coupling'' regime where $\Lambda_N < 300\,{\rm m}$. 
Here the region is showing an example with $\gamma_N = \sqrt{2}$~$(p_N = M_N)$.
For $\gamma_N = \sqrt{1.5}$~($p_N = M_N /2$), the boundary is pushed down by 
factor of two.
For the energies $\gamma_N \gtrsim \sqrt{2}$, 
the total decay is effective only for strong coupling
regime $|\Theta_{e,\mu} |^2 \gtrsim 10^{-4}$, which is already ruled out
by many experiments.
% Eq.~(\ref{Eventmaster4}) is thus well applicable for the following discussions which 
% %treat the region far below the bound shown in Fig.~\ref{tot}.
% focus on the small mixing region 

In Fig.~\ref{tot}, we put in passing the region where the lifetime of the heavy
neutrino becomes long enough so that late time decay of the heavy neutrinos 
may spoil the success of Big Bang Nucleosynthesis (BBN).
Ref.~\cite{BBN,Ruchayskiy:2012si} has studied such a bound for $10\,{\rm MeV} 
< M_N <140 \,{\rm MeV}$ in detail.
For $M_N > 140\,{\rm MeV}$, however, there is no consensus about the constraint 
from BBN.
Here we simply present the region for $\tau > 0.1\, {\rm s}$~\cite{Ruchayskiy:2012si} 
where the heavy neutrinos are not cleared away before the onset of BBN.

%%%%%%%%%%%%%%%%%%%%%%%%%%%%%%%%%%%%%%%%%%%%%%%%%%%%%%%%%%%%%%%%%%%%%%
\subsection{Comparison between T2K and PS191}
%%%%%%%%%%%%%%%%%%%%%%%%%%%%%%%%%%%%%%%%%%%%%%%%%%%%%%%%%%%%%%%%%%%%%%
For each mass eigenstate of the heavy neutrinos, four additional parameters are 
introduced into the Standard Model;
$M_N$ and $\Theta_{e,\mu,\tau}$.
Experiments impose some constrains on the four dimensional space $(M_N, |\Theta_e|, 
|\Theta_\mu|,|\Theta_\tau|)$.
However, the analysis of experimental constraints on the full four-dimensional space 
is a complicated task.
PS191 has made the following assumptions/simplifications in their analysis.
\begin{itemize}
\item Heavy neutrinos are Dirac particles.
\item The NC contributions to the three-body decays of $N$ are neglected.
\item Either $K^+ \to \mu^- N$ or $K^+ \to e^- N$ is dominant in the production.
\end{itemize}
In the following, we first make the same simplifications, just aiming for 
a comparison between T2K and PS191.
In Section~\ref{tau}, we give a comment on the case where the second simplification is
 relaxed so that the three-body decay depends on $\Theta_\tau$.

%%%%%%%%%%%%%%%%%%%%%%%%%%%%%%%%%%%%%%%%%%%%%%%%%%%%%%%%%%%%%%%%%%%%%%
\subsubsection{Two body channels}
%%%%%%%%%%%%%%%%%%%%%%%%%%%%%%%%%%%%%%%%%%%%%%%%%%%%%%%%%%%%%%%%%%%%%%

%%%%%%%%%%%%%%%%%%%%%%%%%%%%%%%%%%%%%%%%%%%%%%%%%%%%%%%%%%%%%%%%%%%%%%
\begin{figure}[t]
  \centerline{
  \includegraphics[scale=0.22]{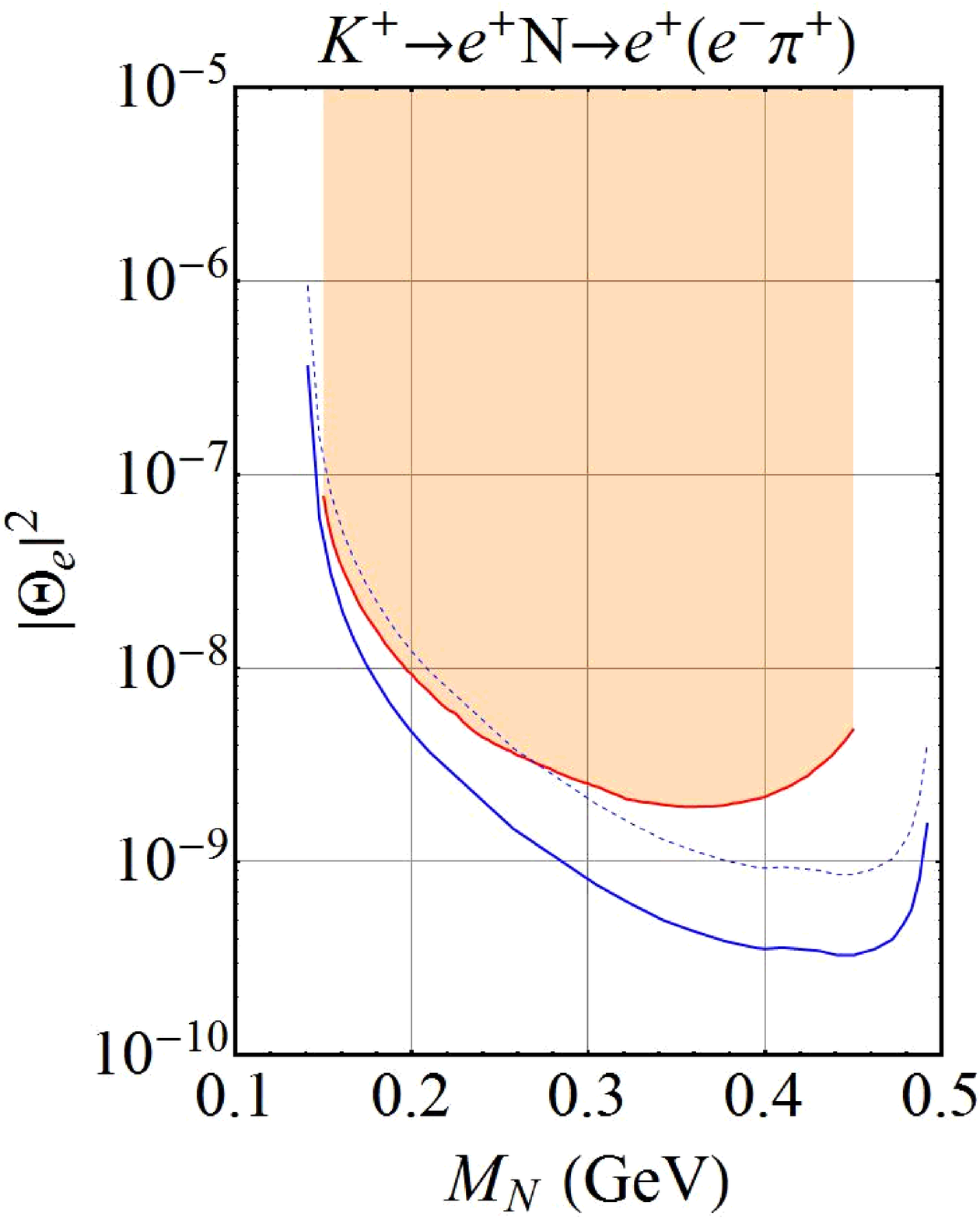}%
  \includegraphics[scale=0.22]{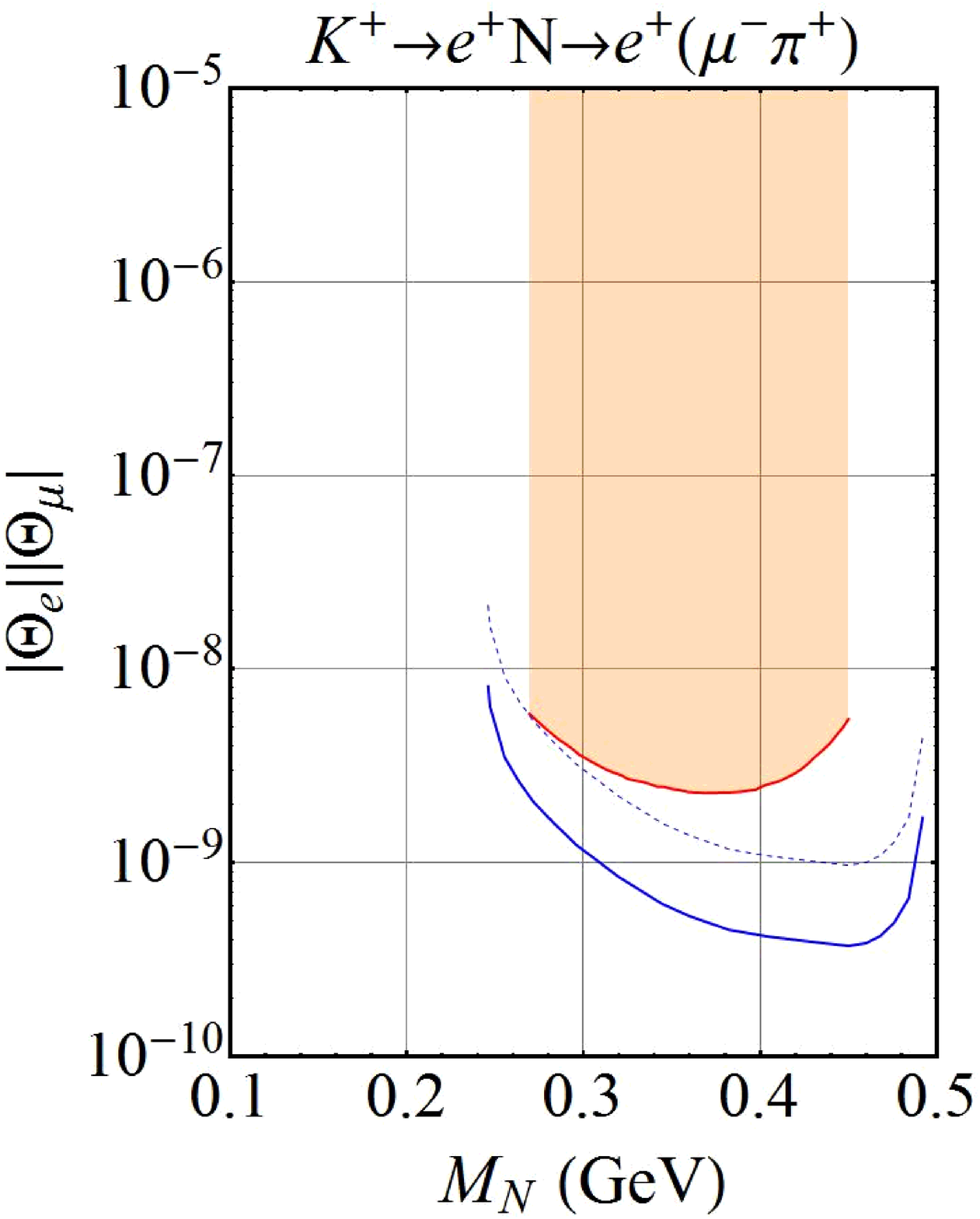}%
  \includegraphics[scale=0.22]{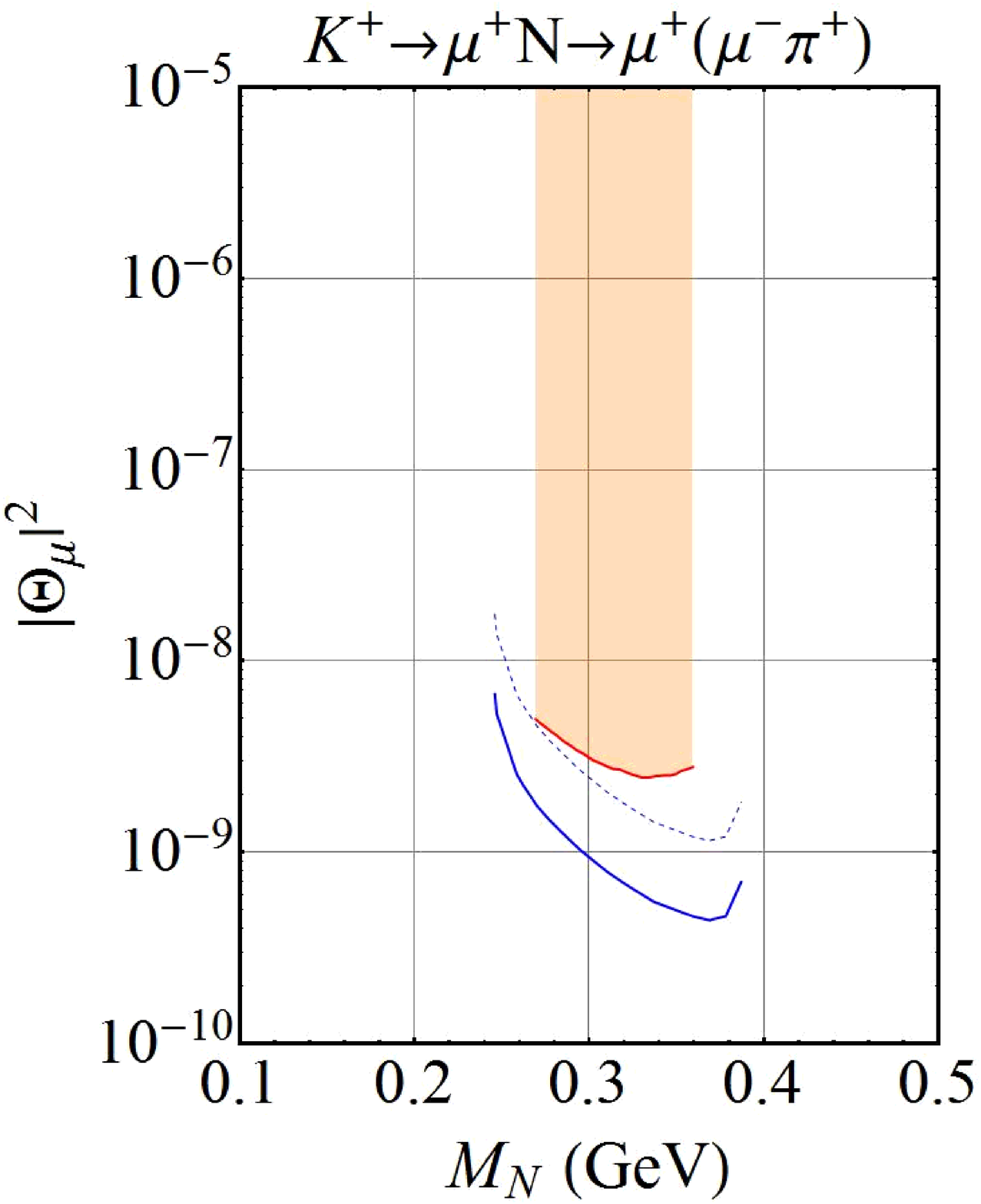}%
  }%
\caption{Expected sensitivity of T2K and upper limits by PS191. 
The blue-solid curves show the 90\% CL upper bounds by T2K at $10^{21}\,{\rm POT}$ 
with the full volume $61.25\,{\rm m^3}$, when no signals are observed.
The blue-dashed curves show the same bounds but with the partial TPC volume
$9.0\,{\rm m^3}$.
The filled regions (with the red-boundary curves) are excluded by 
PS191 at 90\% CL~\cite{PS191}.}
  \label{bounds2}
\end{figure}
%%%%%%%%%%%%%%%%%%%%%%%%%%%%%%%%%%%%%%%%%%%%%%%%%%%%%%%%%%%%%%%%%%%%%% 

Let us first focus on the two-body decays.
Fig.~\ref{bounds2} shows the expected sensitivity for the chains 
$K^+ \to e^+ N \to e^+ (e^- \pi^+)$ (left), $K^+ \to e^+ N \to e^+ (\mu^- \pi^+)$
(middle) and $K^+ \to \mu^+ N \to \mu^+ (\mu^- \pi^+)$ (right), respectively.
In the figures, the red curves show 90\% CL upper bound by PS191~\cite{PS191}.
The blue solid curves show the contour for $2.44$ events, which corresponds to
90\% CL limit when the measured signal and the expected background are 
null~\cite{Feldman:1997qc}.
Here the fiducial volume of the detector is taken as 
$V =3.5 \times 3.5 \times 5.0 = 61.25\,{\rm m^3}$~\cite{ND280}.

The interactions between the active neutrinos and the nuclei in the detector
provide backgrounds for the decay signals $N \to \mu^- \pi^+$ and $N \to e^- \pi^+$.
For instance, the reactions
\begin{eqnarray}
&&\nu_\mu + n \to \mu^- + \pi^+ + n \quad\quad\quad\quad ({\rm CC}- n \pi^+)\nonumber\\
&&\nu_\mu + {}^{16}{\rm O} \to \mu^- + \pi^+ + {}^{16}{\rm O}\quad\quad
({\rm CC- coherent \,\pi^+}) \nonumber
\end{eqnarray}
may become background for $N \to \mu^- \pi^+$.
It is expected that these events account for $4$\% of the whole neutrino events in ND280,
resulting $7300\,{\rm events/10^{21}POT/ton}$~\cite{ND280}.
However the background can be reduced by taking the invariant mass of
$\mu^-$ and $\pi^+$ momenta in the final state.
For the heavy neutrino signal, the event distribution sharply peaks at the heavy 
neutrino mass while the $\nu_\mu$ events provide continuous background.
A serious sensitivity should be estimated together with the invariant mass distribution 
for the $\nu_\mu$ reactions, the energy resolutions, all sort of uncertainties, etc.
Such a thorough analysis is interesting but beyond the scope of this work.
%%%%%%%%%%%%%%%%%%%%%%%%%%%%%%%%%%%%%%%%%%%%%%%%%%%%%%%%%%%%%%%%%%%%%%
\begin{figure}[t]
  \centerline{
  \includegraphics[scale=0.22]{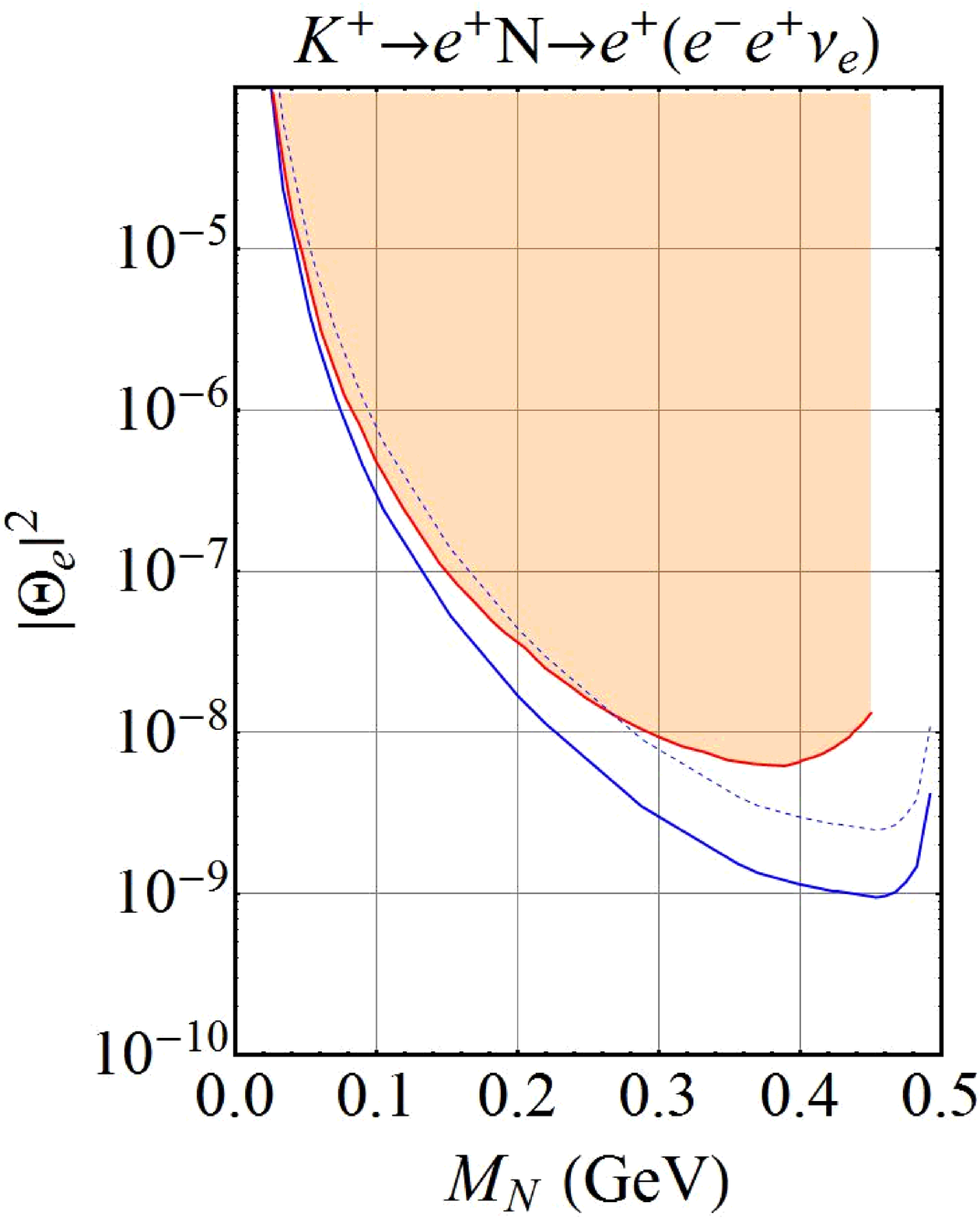}%
  \includegraphics[scale=0.22]{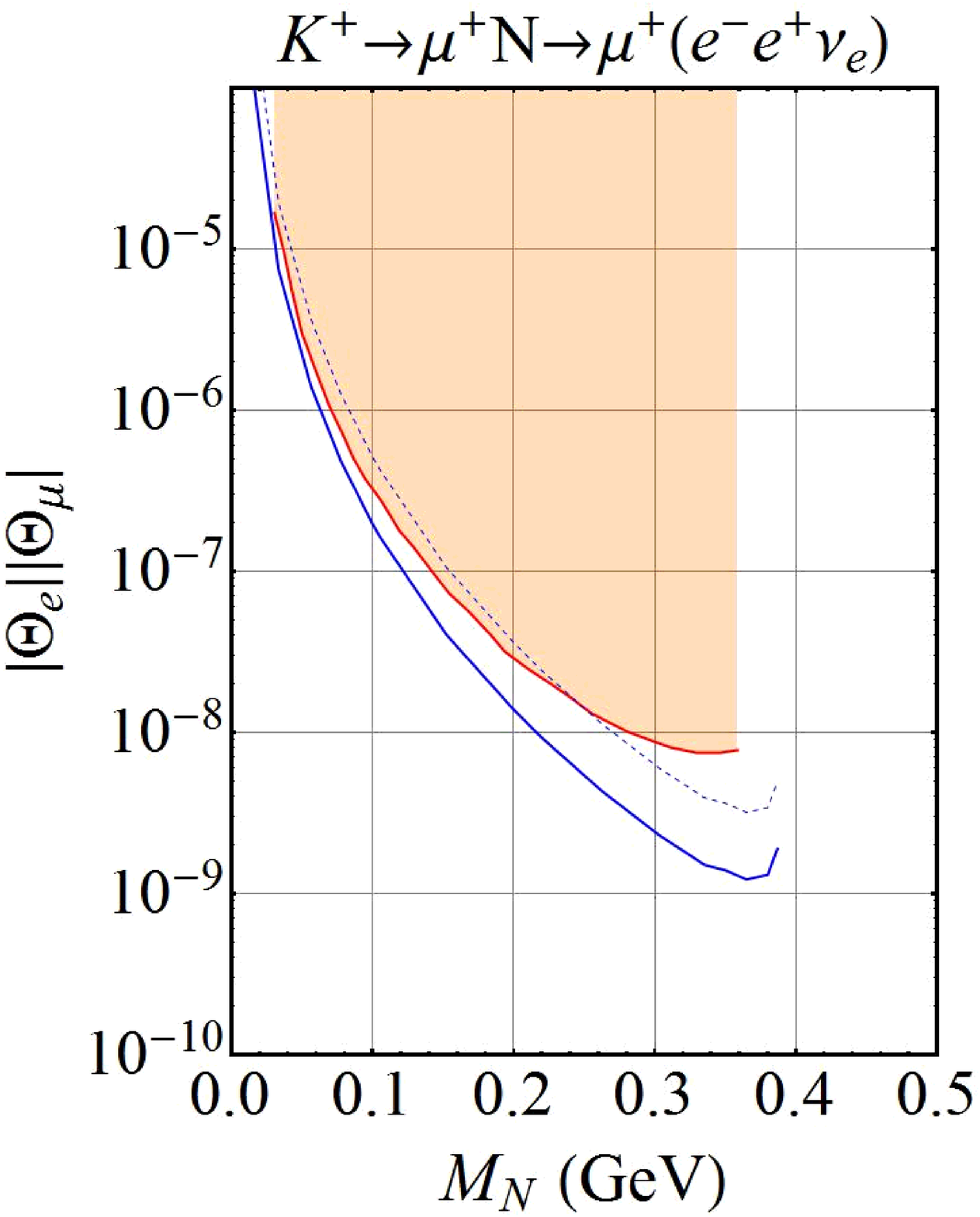}%
  \includegraphics[scale=0.22]{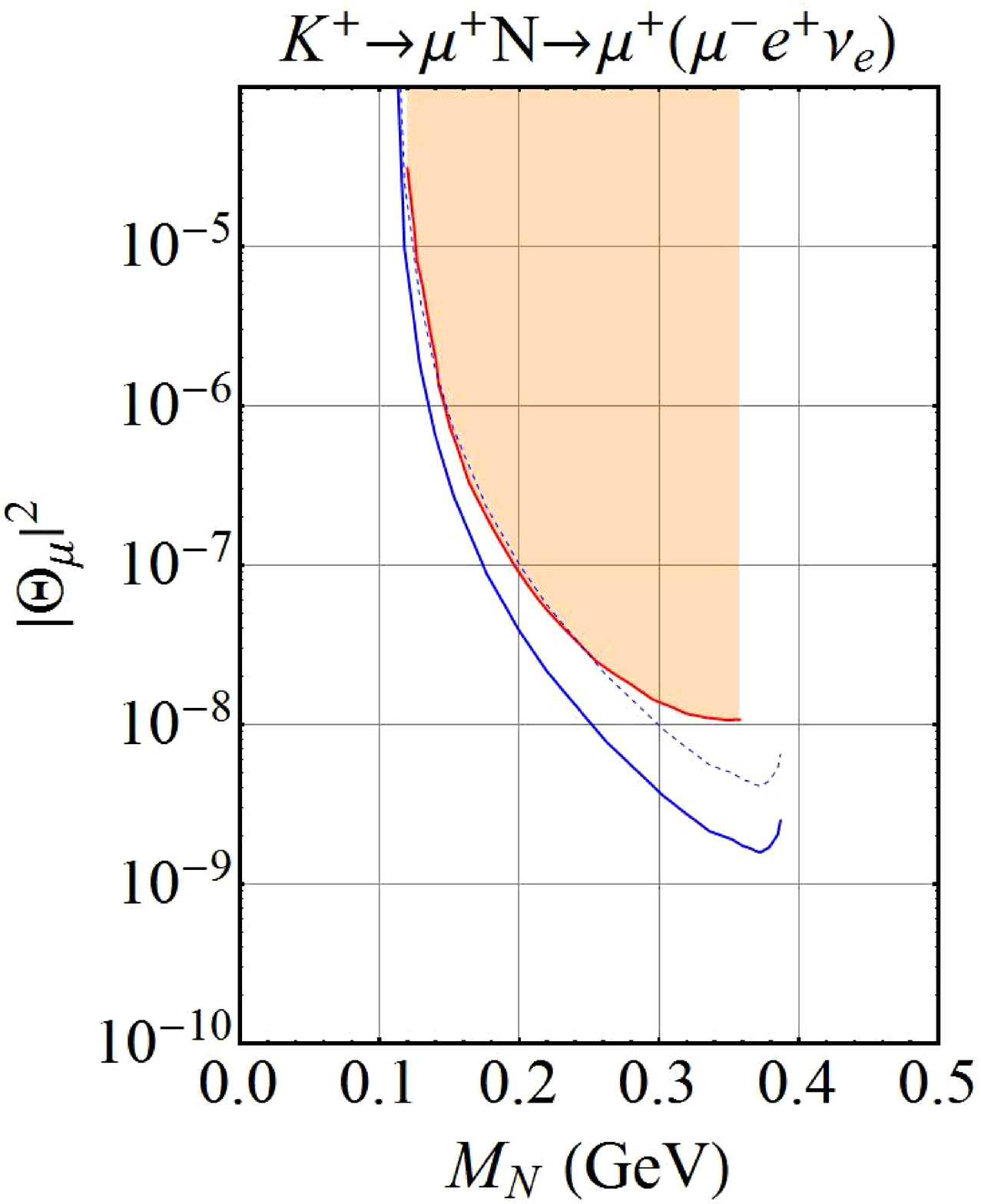}%
  }%
\caption{Same as Fig.~\ref{bounds2} but for the three-body decay channels.}
  \label{bounds3}
\end{figure}
%%%%%%%%%%%%%%%%%%%%%%%%%%%%%%%%%%%%%%%%%%%%%%%%%%%%%%%%%%%%%%%%%%%%%% 

We can further reduce the background by selecting the events taking place
in the TPC volume which is filled by argon gas.
Due to low density of the gas region, the  $\nu_\mu$ events are significantly 
reduced while keeping the signal rates unchanged.
Out of the full volume of $61.25\,{\rm m^3}$, $9.0\,{\rm m^3}$ is filled with the
argon gas~\cite{ND280} and available for this purpose.
According to Ref.~\cite{Karlen}, the total neutrino events taking place in
the argon gas are about $2000$ at $10^{21}\,{\rm POT}$.
Since $4$\% of them become the background, the number of
the background event is expected to be around $80$.
These $80$ events will be further reduced in the bin around the heavy neutrino mass.
Keeping this in mind, we plot the contour for $2.44$ events with $V = 9.0\,{\rm m^3}$
by the dashed curves.

For $N \to e^- \pi^+$, the background processes are produced by the CC
interactions of $\nu_e$.
However, the $\nu_e$ flux is about two orders of magnitude smaller than that of $\nu_\mu$
around the peak energy $\sim 600\,{\rm MeV}$~\cite{flux,D1,Abe:2012av}.
By selecting the events in the gas volume, the background rate is expected to be
less than one for $10^{21}\,{\rm POT}$.
The decay channel $N \to e^- \pi^+$ is more promising than $N \to \mu^- \pi^+$ in
view of signal/background ratio if $|\Theta_e| \sim |\Theta_\mu|$.
%%%%%%%%%%%%%%%%%%%%%%%%%%%%%%%%%%%%%%%%%%%%%%%%%%%%%%%%%%%%%%%%%%%%%%
\subsubsection{Three body channels}
%%%%%%%%%%%%%%%%%%%%%%%%%%%%%%%%%%%%%%%%%%%%%%%%%%%%%%%%%%%%%%%%%%%%%%
Fig.~\ref{bounds3} presents the same plots as Fig.~\ref{bounds2} but 
for the three-body channels;
$K^+ \to e^+ N \to e^+ (e^- e^+ \nu_e)$ (left),
$K^+ \to \mu^+ N \to \mu^+ (e^- e^+ \nu_e)$ (middle),
$K^+ \to \mu^+ N \to \mu^+ (\mu^- e^+ \nu_e)$ (right).

A major obstacle to successful identification of $N \to e^- e^+ \nu$ would be 
$\pi^0$ that is copiously produced by neutrino interactions.
The two photons from $\pi^0$ decay develop to electromagnetic cascades in
the detector material and may mimic the signals.
The subdominant decay mode $\pi^0 \to e^- e^+ \gamma\, (1.17\%)$ may also 
contribute to the background when one of the final-state particle is undetected.
The invariant mass distribution of the electron pair is useful since
it moderately peaks at one-half of the heavy neutrino mass~\cite{ATM}.
The analysis needs anyway precise understanding of the background, and
the detection via $N \to e^- e^+ \nu$ seems less promising than the two-body
modes.

As for $N \to \mu^- e^+ \nu$, the charmed-meson production by the neutrino 
CC interaction~\cite{charm} and successive semi-leptonic decay 
may become the background.
According to Ref.~\cite{charm}, the cross section of the charm production is about 
$1\%$ ($4\%$) of the total CC cross section for $E_\nu = 5\,{\rm GeV}\,(15\,{\rm GeV})$.
Due to the off-axis technic, however, the contributions from such 
high-energy neutrinos are suppressed in the T2K setup.
By selecting the events in argon gas, the neutrino reduction rates can be further
reduced, and $N \to \mu^- e^+ \nu$ may become more or less background free. 

Although PS191 has not studied the signal process $N \to \mu^- \mu^+ \nu$ open for
$M_N > 211\,{\rm MeV}$, it should be emphasized that searching for this di-muon signal 
is also a promising method. 
The main background for $N \to \mu^- \mu^+ \nu$ may be
the charmed-meson production by the neutrino CC interaction~\cite{charm} and 
successive semi-leptonic decay as in the case of $N \to \mu^- e^+ \nu$.
This rate is, however, expected to be small for the neutrino energy in T2K.

%%%%%%%%%%%%%%%%%%%%%%%%%%%%%%%%%%%%%%%%%%%%%%%%%%%%%%%%%%%%%%%%%%%%%%
\subsection{Implications of $\Theta_\tau \neq 0$}
\label{tau}
%%%%%%%%%%%%%%%%%%%%%%%%%%%%%%%%%%%%%%%%%%%%%%%%%%%%%%%%%%%%%%%%%%%%%%

%%%%%%%%%%%%%%%%%%%%%%%%%%%%%%%%%%%%%%%%%%%%%%%%%%%%%%%%%%%%
\begin{figure}[t]
    \begin{center}
\scalebox{0.3}{\includegraphics{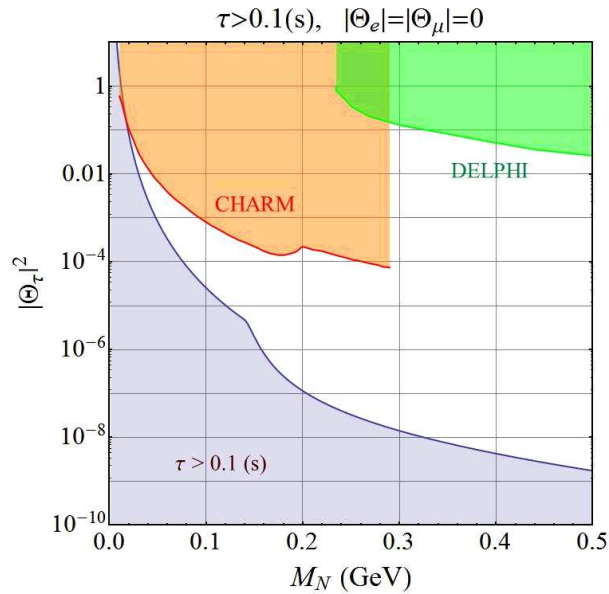}}
\end{center}
\caption{Allowed region of $M_N$-$|\Theta_\tau|^2$ plane.
The upper-filled regions are excluded by CHARM~\cite{Orloff:2002de} and 
DELPHI~\cite{Abreu:1996pa} at 90\% and 95\% CL, respectively.
The lower-filled region is the regime where the lifetime of the heavy neutrino
is longer than $0.1\,{\rm s}$.}
\label{Taubbn}
\end{figure}
%%%%%%%%%%%%%%%%%%%%%%%%%%%%%%%%%%%%%%%%%%%%%%%%%%%%%%%%%%%%
%%%%%%%%%%%%%%%%%%%%%%%%%%%%%%%%%%%%%%%%%%%%%%%%%%%%%%%%%%%%
\begin{figure}[t]
    \begin{center}
\scalebox{0.5}{\includegraphics{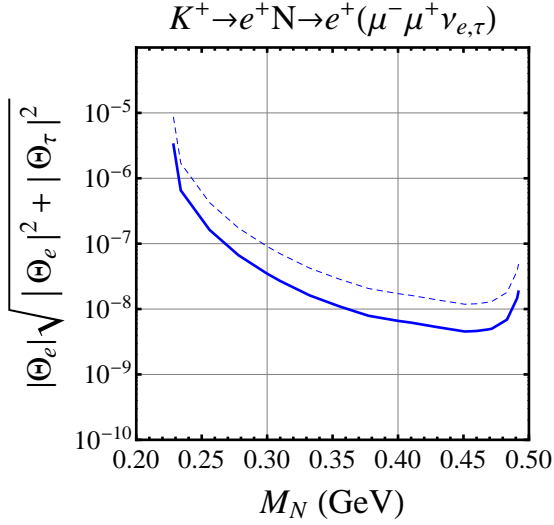}}
\end{center}
\caption{Expected sensitivity for $N \to \mu^- \mu^+ \nu$.
$|\Theta_\mu | =0$ is assumed.}
\label{BoundTtau}
\end{figure}
%%%%%%%%%%%%%%%%%%%%%%%%%%%%%%%%%%%%%%%%%%%%%%%%%%%%%%%%%%%%

So far we have focused on the comparison between T2K and PS191 and
$\Theta_\tau$ is accordingly neglected. 
In this subsection, we comment on several implications of the $\Theta_\tau \neq 0$ case.

Since $\Theta_\tau$ is not involved in the main production processes such as pion
and kaon decays, the experimental constraints of $|\Theta_\tau|^2$ are much wearker
than that of $|\Theta_e|^2$ and $|\Theta_\mu|^2$.
Fig.~\ref{Taubbn} shows the allowed region of $M_N$-$|\Theta_\tau|^2$ plane.
The upper-filled regions are excluded by CHARM~\cite{Orloff:2002de} and 
DELPHI~\cite{Abreu:1996pa} at 90\% and 95\% CL, respectively.
The lower-filled region is the regime where the lifetime of the heavy neutrino
is longer than $0.1\,{\rm s}$.
By comparing Fig.~\ref{Taubbn} and Fig.~\ref{tot}, it is seen that,
unlike  $|\Theta_{e,\mu}|^2$, $|\Theta_\tau|^2$ can be large enough so that the BBN 
constraint is avoided without contradicting with the upper bound from the direct 
experiments.

While $\Theta_\tau$ is not involved in the production processes, the detection 
processes in general depend on $\Theta_\tau$ since $N \to e^- e^+ \nu$ and 
$N \to \mu^- \mu^+ \nu$ are induced not only by the CC but also by 
the NC interactions~\cite{Kusenko:2004qc}.
This means the experimental setup discussed in this paper has sensitivities
to certain combinations of $|\Theta_{e,\mu}|^2$ and $|\Theta_\tau|^2$.
To demonstrate this, let us focus on a simple case where $|\Theta_e|^2, |\Theta_\tau|^2
\gg |\Theta_\mu|^2$ and $M_N > 211\,{\rm MeV}$.
In this case, the heavy neutrinos are produced by $K^+ \to e^+ N$ and
can be detected by $N \to \mu^- \mu^+ \nu$.
Since $|\Theta_\mu|^2$ is small, the decay process $N \to \mu^- \mu^+ \nu$ is conducted 
only by the NC interactions and the decay width becomes proportional to $|\Theta_e |^2 
+ |\Theta_\tau |^2$.
Therefore by analyzing the dimuon signals in the detector, one is able to constraint
$|\Theta_e |\sqrt{|\Theta_e |^2 + |\Theta_\tau |^2}$ for $M_N > 211\,{\rm MeV}$ 
(or discover the heavy neutrino).

Fig.~\ref{BoundTtau} shows the expected sensitivity for $N \to \mu^- \mu^+ \nu$.
As in Fig.~\ref{bounds2} and Fig.~\ref{bounds3}, the solid (dashed) curve is
the contour for $2.44$ events with $V =61.25\,\, (9.0)\,{\rm m^3}$.
From Fig.~\ref{BoundTtau} and Fig.~\ref{Taubbn}, one can see that there exists the 
parameter regime where the dimuon signal is sizable 
while the success of BBN is unspoiled.
For example, it is seen that 
the combination $|\Theta_e | = 10^{-4.5}$ and $|\Theta_\tau|^{-2.5}$ is BBN safe but 
$\mathcal{O}(10)-\mathcal{O}(10^2)$ events of $N \to \mu^- \mu^+ \nu$ are expected
for $300\,{\rm MeV} \lesssim M_N \lesssim 400\,{\rm MeV}$.
Interestingly, this case also predicts $\mathcal{O}(1)-\mathcal{O}(10)$ events of
$N \to e^-\pi^+$ (see the left pannel of Fig.~\ref{bounds2}), so that the heavy 
neutrino model can be tested in multi-dimensional way.

%%%%%%%%%%%%%%%%%%%%%%%%%%%%%%%%%%%%%%%%%%%%%%%%%%%%%%%%%%%%%%%%%%%%%%
\section{Conclusions}
\label{conclusion}
%%%%%%%%%%%%%%%%%%%%%%%%%%%%%%%%%%%%%%%%%%%%%%%%%%%%%%%%%%%%%%%%%%%%%%
In this paper, we have focused on the heavy (sterile) neutrinos produced by
kaon decays and explored the feasibility of their detection at 
the existing facilities of the accelerator-based neutrino experiment.
Taking the T2K experiment as a typical example, we have estimated the
heavy neutrino fluxes produced in the beam line and calculated the 
event rates of their decay taking place inside the detector.

Due to massive nature of the heavy neutrino, the spectrum of the heavy neutrino is  
significantly different from that of the ordinary neutrinos.
The ordinary neutrinos are emitted to various directions in the laboratory
frame due to their tiny masses. 
On the other hand the heavy neutrinos carrying a large mass tend to be emitted to 
the forward directions and frequently hit the detector.
This is a unique advantage of the experiments in which the parent mesons decay in flight
with sufficient gamma factors.

Among various decay modes, $N \to e^- \pi^+$ open for $M_N > 140\,{\rm MeV}$
is one of the most promising channels for detection because of its larger rate 
and lower background.
The backgrounds from the active neutrino reactions can be reduced
by selecting the events occurring in the regions filled with no material.
In the T2K near dector, the TPC volume $9\,{\rm m^3}$ filled with argon gas plays
this role.
The expected sensitivity for this mode is better than that of PS191,
which has placed the most stringent bound on the heavy neutrino mixing.

The three body modes $N \to e^- e^+ \nu$, $N \to \mu^- e^+ \nu$, 
and $N \to \mu^- \mu^+ \nu$ are also interesting signals to search for.
In particular, $N \to e^- e^+ \nu$ and $N \to \mu^- \mu^+ \nu$ are conducted
not only by the charged current but also the neutral current, so that the tau
flavor mixing $\Theta_\tau$ is involved in the detection probabilities.
Since $|\Theta_\tau|$ is less constrained than $|\Theta_e|$ and $|\Theta_\mu|$, 
the above two modes are not necessarily suppressed when $|\Theta_e|$ and $|\Theta_\mu|$
are small such that the two-body modes $N \to e^- \pi^+$ and $N \to \mu^- \pi^+$
are beyond the reach.

Finally, we would like to emphasize that two quasi-degenerate heavy neutrinos of 
$\mathcal{O}(100)\,{\rm MeV} - \mathcal{O}(10)\,{\rm GeV}$ can account for
not only the neutrino masses in oscillation experiments but also
the baryon asymmetry of the universe~\cite{BAU,Asaka:2005pn,BAU2}. 
The heavy neutrinos studied in this work are thus quite interesting targets to
search for.
Furthermore, Ref.~\cite{Fuller:2009zz} reports that the sterile
neutrinos with mass $\sim 200\,{\rm MeV}$ could facilitate the energy transport
from the supernova core to the schock front, prompting a successful explosion.
The needed mixing is either $|\Theta_\tau|^2 > 10^{-8}$ or 
$10^{-7}-10^{-8}$ for $|\Theta_\mu|^2$.
Interestingly, T2K can probe latter case via the $N \to \mu^- e^+ \nu$ mode
(see the right pannel of Fig.~\ref{bounds3}).
In addition, heavy neutrinos with masses smaller than ${\cal
O}(100)$ MeV may give a significant effect on the neutrinoless double beta 
decays~\cite{AEI}.
According to a rough estimation in Table~\ref{t1}, MiniBooNE and SciBooNE have 
comparable abilities to T2K so that these experiments equally have the 
chance to probe these interesting possibilities.
Serious analyses by these collaborations may lead to the discovery of the heavy 
neutrinos and revolutionize neutrino physics.

%%%%%%%%%%%%%%%%%%%%%%%%%%%%%%%%%%%%%%%%%%%%%%%%%%%%%%%%%%%%%%%%%%%%%%
\subsection*{Acknowledgments}
This work is supported by the Young Researcher Overseas Visits 
Program for Vitalizing Brain Circulation Japanese in JSPS (No.~R2209).
T.A. is supported by KAKENHI (No. 21540260) in JSPS.
A.W. would like to thank E.~K.~Akhmedov and T.~Schwetz for useful discussions.
We would like to thank Particle and Astroparticle Division of
Max-Planck-Institut f\"ur Kernphysik at Heidelberg for hospitality.
\bigskip
%%%%%%%%%%%%%%%%%%%%%%%%%%%%%%%%%%%%%%%%%%%%%%%%%%%%%%%%%%%%%%%%%%%%%%

\appendix
%%%%%%%%%%%%%%%%%%%%%%%%%%%%%%%%%%%%%%%%%%%%%%%%%%%%%%%%%%%%%%%%%%%%%%
\section{Details on the flux calculation}
\label{fluxculc}
%%%%%%%%%%%%%%%%%%%%%%%%%%%%%%%%%%%%%%%%%%%%%%%%%%%%%%%%%%%%%%%%%%%%%%

\begin{figure}[t]
\begin{center}
\scalebox{0.3}{\includegraphics{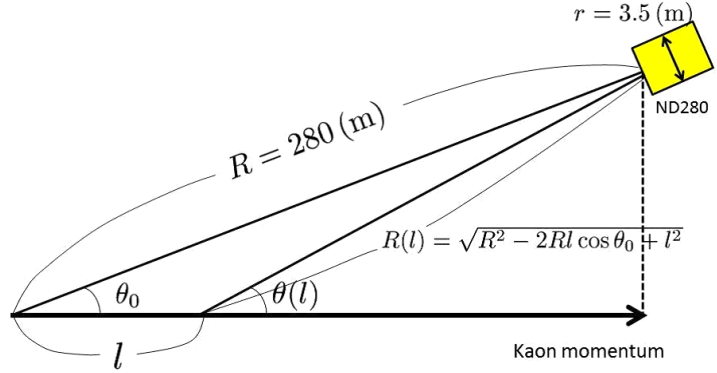}}
\end{center}
\caption{Geometry of the near detector ND280 and the kaon momentum.}
\label{geo}
\end{figure}

In this appendix, we present technical details on the flux calculation discussed
in Section~\ref{flux}. 
Let us start from the source term Eq.~(\ref{Snu}),
\begin{eqnarray}
S_\nu(E_\nu, \theta, \phi, l) &=& 
\int_0^\infty \!\!\!dp_K \,\phi_K(p_K,l)\left( \frac{m_K}{p_K} \right)
\frac{d^3 \Gamma}{dE_\nu d\cos\theta d\phi}.
\label{Snu2}
\end{eqnarray}
As discussed in Section~\ref{flux}, the neutrino flux $\phi_{\nu_\mu}(E_\nu)$
of Eq.~(\ref{phi2}) is obtained by integrating this over $\theta,\phi,l$.
The computation is greatly simplified if the kaon momentum is parallel to the beam axis.
The detector geometry for this case is presented in Fig.~\ref{geo}.
Under this simplification, the neutrino flux follows
\begin{eqnarray}
\phi_{\nu_\mu}(E_\nu) \,=\, \frac{\Delta \phi}{A}
\int_0^{l_f}\!\!\! dl \int_{-1}^{1} \!\!d\!\cos\theta 
\,S_\nu(E_\nu, \theta, l)\,P'(\theta,\theta_0),
\label{phi22}
\end{eqnarray}
where $\Delta \phi = r/R\sin\theta_0$ and $l_f = 96\,{\rm m}$.
The function $P'(\theta,\theta_0)$ projects out the general polar angles into the ones
with which $\nu_\mu$ pass through the near detector.
It is given by
\begin{eqnarray}
P'(\theta,\theta_0)
= H(\theta - \theta^-)H(\theta^+ - \theta),
\end{eqnarray}
where $H(x)$ is the Heaviside step function and
\begin{gather}
\theta^- = \theta(l) -\frac{r}{2R(l)}\,,\quad\quad
\theta^+ = \theta(l) +\frac{r}{2R(l)},
\end{gather}
\begin{eqnarray}
\theta(l) = \arcsin\left[ \,\frac{R\sin\theta_0}{R(l)} \, \right],
\quad
R(l) = \sqrt{R^2 - 2Rl \cos\theta_0 + l^2},
\end{eqnarray}
with $R = 280\,{\rm m}$ and $r = 3.5\,{\rm m}$.
%%%%%%%%%%%%%%%%%%%%%%%%%%%%%%%%%%%%%%%%%%%%%%%%%%%%%%%%%%%%
%%%%%%%%%%%%%%%%%%%%%%%%%%%%%%%%%%%%%%%%%%%%%%%%%%%%%%%%%%%% 

\begin{figure}[t]
\begin{center}
\scalebox{0.4}{\includegraphics{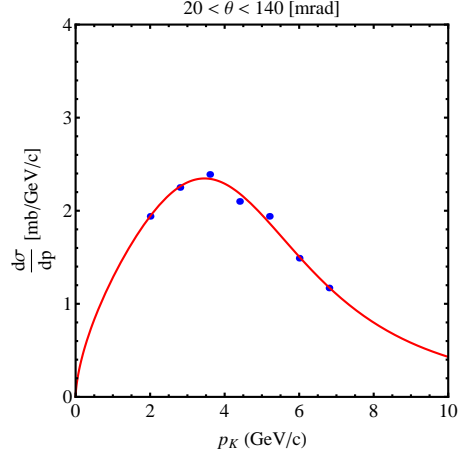}}
\end{center}
\caption{The $K^+$ spectrum $d\sigma/dp_K$ in proton--carbon collisions at
$31\,{\rm GeV}$ proton energy.
The dots are the best-fit values of the data~\cite{Abgrall:2011ae} 
and the solid line is the fit.}
\label{dsdE}
\end{figure}

The $K^+$ spectrum $d\sigma/dp_K$ in the proton collision with a graphite target 
is measured by NA61/SHINE Collaboration~\cite{Abgrall:2011ae}.
Fig.~\ref{dsdE} shows the best-fit values of the data~\cite{Abgrall:2011ae} and 
an example of the fit.
The spectrum is the average over the $K^+$ polar angles (relative to the beam axis)
$20-140\, {\rm mrad}$ where the probability that daughter neutrinos
pass through ND280 is high~\cite{Abe:2012av,Abgrall:2011ae}.
We assume that the shape of the kaon spectrum $\phi_K(p_K)$ is 
not far from this measured spectrum and use the following fitting;
\begin{gather}
\phi_K(p_K) \,=\, a_0 \!\left(\frac{d\sigma'}{dp_K}\right),\nonumber\\
\frac{d\sigma'}{dp_K} = \frac{\phi_L \phi_H}{\phi_L + \phi_H},\quad\quad
\phi_L = a_L \,{p_K}^{b_L}, \quad \phi_H = a_H (p_K + p_0)^{-b_H}, 
\end{gather}
where $a_L = 1.285\,\,{\rm mb\, GeV}^{-(b_1 + 1)}$, $b_L = 0.6289$,
$a_H = 793.5\,\,{\rm mb\, GeV}^{-(b_2 + 1)}$, $b_H = 3.230$, with which 
$d\sigma'/dp_K = d\sigma/dp_K$ at $p_0 =0$. 
A positive value of $p_0$ shifts the peak of $d\sigma/dp_K$ to lower energies.
%%%%%%%%%%%%%%%%%%%%%%%%%%%%%%%%%%%%%%%%%%%%%%%%%%%%%%%%%%%%
%%%%%%%%%%%%%%%%%%%%%%%%%%%%%%%%%%%%%%%%%%%%%%%%%%%%%%%%%%%%

The differential decay width in Eq.~(\ref{Snu2}) is given by
\begin{eqnarray}
\frac{d^3\Gamma}{dE_\nu d\!\cos\theta d\phi} = \frac{M^2}{2E_K}
\frac{E_\nu}{8\pi^2}
\,\delta\Big[f(p_K) \Big],
\end{eqnarray}
where 
\begin{eqnarray}
&&M^2 \,\equiv\,
2G_F^2 f_K^2 m_K^4 |V_{us}|^2 \left[ \left(\frac{m_\mu}{m_K}\right)^2 - 
\left(\frac{m_\mu}{m_K}\right)^4 \right],\\
&&f(p_K) \,=\,m_K^2 - m_\mu^2 -2E_\nu\sqrt{p_K^2 + m_K^2} + 2p_K E_\nu\cos\theta.
\end{eqnarray}
An explicite expression of Eq.~(\ref{phi22}) is written as
\begin{eqnarray}
\phi_{\nu_\mu}(E_\nu) \,=\, 
\frac{\Delta \phi}{A}
\int_0^{l_f}\!\!\! dl \int_{-1}^{1} \!\!d\!\cos\theta 
\,F_{\rm low}\,P'(\theta,\theta_0),
\label{phiL}
\end{eqnarray}
for $0< E_\nu \leq (m_K^2 - m_\mu^2)/2m_K$
and 
\begin{eqnarray}
\phi_{\nu_\mu}(E_\nu) \,=\, 
\frac{\Delta \phi}{A}
\int_0^{l_f}\!\!\! dl \int_{\cos\theta_c}^{1} \!\!d\!\cos\theta 
\,F_{\rm high}\,P'(\theta,\theta_0)
\label{phiH}
\end{eqnarray}
for $E_\nu \geq (m_K^2 - m_\mu^2)/2m_K$.
Here 
\begin{eqnarray}
F_{\rm low} &\,=\,&
 \frac{M^2}{8\pi^2}
\left[\, \phi(p_K^+,l)\left( \frac{m_K}{p_K^+} \right)
\frac{1}{2E_K(p_K^+)}\frac{1}{2\left |\cos\theta - 
\frac{p_K^+}{E_K(p_K^+)}
\right|}\,\, 
\right],\\
F_{\rm high}
 &=& \frac{M^2}{8\pi^2}
\left[\,
\phi(p_K^-,l)\left( \frac{m_K}{p_K^-} \right)
\frac{1}{2E_K(p_K^-)}\frac{1}{2\left|\cos\theta - 
\frac{p_K^-}{E_K(p_K^-)}
\right|} \right. \nonumber\\
&&\left.
\quad\quad\quad\quad\,+\,\,\phi(p_K^+,l)\left( \frac{m_K}{p_K^+} \right)
\frac{1}{2E_K(p_K^+)}\frac{1}{2\left|\cos\theta - 
\frac{p_K^+}{E_K(p_K^+)}
\right|}\,\, 
\right],
\end{eqnarray}
\begin{gather}
p_K^\pm \,=\,
\frac{(m_K^2 -m_\mu^2)\cos\theta \pm \sqrt{(m_K^2 -m_\mu^2)^2 -4(1-\cos^2\theta)m_K^2
E_\nu^2}}{2(1-\cos^2\theta)E_\nu},\\
\theta_c = \arcsin\left[\, \frac{m_K^2 - m_\mu^2}{2m_K E_\nu} \right].
\end{gather}

For the heavy neutrino production $K^+ \to \mu^+ N$, the differential decay width is 
given by
\begin{eqnarray}
\frac{d^3\Gamma_1}{dE_N d\!\cos\theta d\phi} = \frac{|\Theta_\mu|^2M_1^2}{2E_K}
\frac{p_N}{8\pi^2}
\,\delta\Big[g(p_K) \Big],
\label{dGam1}
\end{eqnarray}
where 
\begin{eqnarray}
&&M_1^2 \,\equiv\,
2G_F^2 f_K^2 m_K^4 |V_{us}|^2 \left[ \left(\frac{M_N}{m_K}\right)^2 
+ \left(\frac{m_\mu}{m_K}\right)^2
- \left( \left(\frac{M_N}{m_K}\right)^2 - \left(\frac{m_\mu}{m_K}\right)^2  \right)^2 
\right],\\
&&g(p_K) \,=\,m_K^2 +M_N^2 - m_\mu^2 -2E_N\sqrt{p_K^2 + m_K^2} + 2p_K p_N\cos\theta.
\end{eqnarray}
The formula for $K^+ \to e^+ N$ is obtained by making the replacement $\mu \to e$
in the above.
By replacing the differential decay width in Eq.~(\ref{Snu2}) with Eq.~(\ref{dGam1}),
it is straightforward to obtain the formulas for $\phi_N(E_N)$ similar to 
Eq.~(\ref{phiL}) and Eq.~(\ref{phiH}).

%%%%%%%%%%%%%%%%%%%%%%%%%%%%%%%%%%%%%%%%%%%%%%%%%%%%%%%%%%%%%%%%

\end{document}